\documentclass[aps,amsmath,amssymb,prb,reprint,showpacs]{revtex4-1}
\usepackage{graphicx,subfigure,color}

\newcommand{\bs}{\boldsymbol}
\newcommand{\ua}{\uparrow}
\newcommand{\da}{\downarrow}
\newcommand{\uda}{\updownarrow}
\newcommand{\lra}{\leftrightarrow}

\newcommand{\bfL}{\mathbf{L}}
\newcommand{\bfT}{\mathbf{T}}
\newcommand{\bfX}{\mathbf{X}}
\newcommand{\bfW}{\mathbf{W}}
\newcommand{\bfr}{\mathbf{r}}
\newcommand{\bft}{\mathbf{t}}
\newcommand{\bfs}{\mathbf{s}}
\newcommand{\bfp}{\mathbf{p}}

\newcommand{\bfI}{\mathbf{I}}
\newcommand{\bfJ}{\mathbf{J}}
\newcommand{\bfj}{\mathbf{j}}
\newcommand{\bfK}{\mathbf{K}}
\newcommand{\bfk}{\mathbf{k}}
\newcommand{\bfe}{\mathbf{e}}
\newcommand{\bfalp}{\boldsymbol{\alpha}}
\newcommand{\bfbet}{\boldsymbol{\beta}}
\newcommand{\bfgam}{\boldsymbol{\gamma}}
\newcommand{\bftau}{\boldsymbol{\tau}}
\newcommand{\bfsig}{\boldsymbol{\sigma}}
\newcommand{\bfSig}{\boldsymbol{\Sigma}}
\newcommand{\diag}{\text{\bf diag}}
\newcommand{\tr}{\text{\bf tr}}

\newcommand{\orbA}{\overrightarrow{\mathbf{a}}}
\newcommand{\orbB}{\overrightarrow{\mathbf{b}}}
\newcommand{\orj}{\overrightarrow{j}}
\newcommand{\orbalp}{\overrightarrow{\bfalp}}
\newcommand{\orbbet}{\overrightarrow{\bfbet}}
\newcommand{\orbPsi}{\overrightarrow{\bs{\Psi}}}

\newcommand{\olbA}{\overleftarrow{\mathbf{a}}}
\newcommand{\olbB}{\overleftarrow{\mathbf{b}}}
\newcommand{\olj}{\overleftarrow{j}}
\newcommand{\olbalp}{\overleftarrow{\bfalp}}
\newcommand{\olbbet}{\overleftarrow{\bfbet}}
\newcommand{\olbPsi}{\overleftarrow{\bs{\Psi}}}

\newcommand{\olrbA}{\overleftrightarrow{\mathbf{a}}}
\newcommand{\olrbB}{\overleftrightarrow{\mathbf{b}}}
\newcommand{\olrj}{\overleftrightarrow{j}}

\newcommand{\bkp}{\boldsymbol{k}_{\parallel}}

\newcommand{\kpar}{k_{\parallel}}

\newcommand{\bfrprime}{\mathbf{r}'}

\newcommand{\oscr}[1]{\overrightarrow{#1}^{\sim}}
 
\begin{document}
\title{Application of the Landauer formalism to the calculation of spin current}

\author{V. Fadeev and A. Umerski}
\affiliation{Department of Mathematics and Statistics, Open University, Milton Keynes MK7 6AA,
U.K.}

\date{\today}

\begin{abstract}
In this communication we apply the Landauer method and transfer matrix formalism to the calculation of spin current in magnetic multilayered structures within a ballistic quantum-mechanical regime.
The method provides an elegant and intuitive formalism with which to study spin current properties and within which closed-form expressions with a transparent physical interpretation can be obtained.
We apply the method to illuminate origin and the symmetry properties of the various spin current components within the parabolic band approximation.
We also apply the stationary phase approximation to develop asymptotic approximations to the total spin current within this formalism, and show that these give excellent agreement with full numerical calculations for both barrier and well systems.
\end{abstract}


\maketitle

\section{Introduction}
The study of electron (hole) spin-transport in magnetic nanoscopic magnetic multilayers has been of great interest in the past few decades giving rise to the developing field of spintronics 
\cite{bandyopadhyay2008introduction, RevModPhys.76.323}.
Among the first spintronic effects observed were oscillatory interlayer exchange coupling (IEC) \cite{Parkin90}, giant magnetoresistance (GMR)  \cite{Grunberg86,Baibich88} and tunneling magnetoresistance (TMR) \cite{Mathon01,Butler01,Parkin04,Yuasa04}.
More recently a great deal of effort has been devoted to the study of spin-transfer torque (STT) \cite{Slonczewski1996, Slonczewski1999, Slonczewski2002, Stiles2006, Edwards2006273}:  a mechanism by which a spin-polarised current can be used to effect switching of magnetic moments. 
This topic is of particular interest because of its potential application to magnetic random access memory  \cite{Butler2004,Akerman2005} -- a principal contender for the next generation of memory devices.

All these effects can be described by a single underlying physical mechanism, namely the flow of spin current and its interaction with the potentials and magnetic moments of its environment.
Initially these phenomena were described in terms of simple parabolic band models in which wave functions were matched across interfaces \cite{Slonczewski1996, Edwards:2007}. 
However various approximations such as, infinite exchange splitting in the magnetic components of the multilayer and/or, exact matching of the potentials of one spin band across the entire multilayer, are often required in order to yield tractable analytic results.
Although physically transparent, this approach is limited to toy models.
Nowadays the Keldysh formalism \cite{Keldysh1964,Caroli1971}, which is designed to calculate the transport properties of a non-equilibrium system system in its steady state, is the principle technique for calculating spin currents.
The advantage of this approach is that it can be implemented to a fully realistic band structure, but at the cost of lacking physical transparency and only being tractable in hefty numerical calculations.

A third approach is to use the method of Landauer within the transfer-matrix formalism, which is widely applied to the calculation of charge current \cite{Landauer1988,Stone1988}.  
The advantage of this formalism is that it provides an elegant and physically transparent description of spin currents which allows closed form expressions to be obtained in many circumstances and also allows us to explore spin current structure and symmetry properties.
Additionally, like the Keldysh formalism, the Landauer method allows us to separate the bias dependent and independent parts of the spin current.
In principle it may also be possible to generalise the approach to multi-orbital models when one expresses the transfer matrices in terms of Greens' functions (see eg. Appendix H of ref.\cite{Economou2006}).
  Such an extension may well alleviate some of the numerical difficulties (sharp spikes in the spin current density) which occur in Keldysh formalism calculations of realistic metallic multilayer systems in the in the ballistic limit.

The extension of the Landauer formalism to deal with spin currents has been briefly discussed in a few previous communications  \cite{Waintal2000,Edwards2006273,Ralph2008,Human2014}.
However, these publications either concentrated: on toy models with simplifying assumptions, or the diffusive  
regime, or dealt entirely with numerical calculations.
In this communication we develop the Landauer formalism within the ballistic limit in more detail than previously considered.
 In particular, we apply it to understanding the origin, symmetry and asymptotic properties of the components of the spin current for a standard polarizing-magnet/non-magnetic spacer/switching-magnet geometry depicted in Fig.~\ref{fig2}.
 We deduce several results concerning the spin current analytically and relatively straightforwardly and within this single unified framework. Some of these results have previously been observed through numerical studies within the Keldysh formalism, so their deduction within the Landauer formalism provides a reassuring check that the physics is consistent across both frameworks.  
 Finally we show how the Landauer formalism can be used to obtain accurate analytic approximations to the $k$-space integrated spin current for both quantum well and barrier profiles.

This manuscript is organised as follows.  In sections~\ref{sec:back} and \ref{sec:land} we develop the Landauer method within the transfer matrix formalism for a simple one electron parabolic band model.  Section~\ref{sec:origin} applies this construct to demonstrate that the out-of-plane spin current in the spacer arises as a consequence of reflections of carriers between the polarizing and switching magnets.   
Although this has been explored previously in special cases (e.g. see Ref.\cite{Stiles2002,Ralph2008} and references therein), we include it here both for completeness and because some of the results will be required in the remainder of the paper which has not been previously discussed.
Section~\ref{sec:symmetry} examines the symmetry properties of the spin current and explains why the bias dependent part of the out-of-plane spin current vanishes in the spacer if the multilayer is symmetric or there is exact matching of potentials of one spin band across the multilayer -- effects which have previously been observed only numerically \cite{Edwards2005,Tang2010}.
In addition we show that, like the charge current, the in-plane components of the left- and right-moving spin currents are equal and opposite, which explains mathematically why only the out-of-plane component survives in the absence of bias, giving rise to the oscillatory interlayer exchange coupling.
Finally, in Section \ref{sec:asymptotic} we apply the stationary phase approximation to the calculation of total spin current, accurate in the limit of relatively weak reflection and large spacer thickness. Such approximations allow us to understand the behaviour of the spin current and could be of considerable practical use if extended to multi-orbital models where accurate numerical calculation of spin current is very challenging.

\section{Background}\label{sec:back}
We consider an electron (carrier) moving in the $y$ direction through a magnetic material. The spin quantization axis is assumed to be in the $z$ direction, and the magnetic moment is rotated in the $xz$-plane by an angle $\theta$ with respect to the $z$-axis.
The spin-$\tfrac12$ Schr\"{o}dinger equation 
then has potential
\[
 {\bf v}(\theta) = {\bf s}(\theta)^{-1}{\bf v}(0){\bf s}(\theta)\]
 where the exchange field ${\bf v}(0)$ and rotation matrix ${\bf s}(\theta)$ are given by
\[{\bf v}(0)
=v_0 \mathbf{1} + \frac{\Delta}{2} \boldsymbol{\sigma}_z,\qquad 
{\bf s}(\theta) = e^{i \bs{\sigma}_y \theta/2}.\]
where $\mathbf{1}$ is the $2 \times2$ unit matrix and $\bs{\sigma}_y$ and $\bs{\sigma}_z$ are the respective Pauli matrices. 

In a magnetic medium, the Schr\"{o}dinger has (time-independent) eigensolutions 
 \[ \bs{\psi}={\bf s}(\theta)^{-1}\bs{\phi} \quad \text{where} \quad  
 \bs{\phi} =
 \begin{bmatrix}
\alpha^{\uparrow} e^{i k^{\uparrow} y}\\ \alpha^{\downarrow} e^{i k^{\downarrow} y}
\end{bmatrix} +
 \begin{bmatrix}
\beta^{\uparrow} e^{-i k^{\uparrow} y}\\ \beta^{\downarrow} e^{-i k^{\downarrow} y}
\end{bmatrix},
 \]
where  $k^{\updownarrow}=\sqrt{2m(E-v_0\mp\Delta)/\hbar^2-k_x^2-k_z^2}$, is the $k$-vector in the $y$-direction. In a non-magnetic medium, $\Delta=0$ and $k^{\updownarrow}=k$, so the wavefunction is given by
\begin{equation}
\bs{\phi}=
  e^{i k y} \bfalp +
 e^{-i k y} \bfbet,
\label{eq:wf1}
\end{equation}
 where $\bfalp=\begin{bmatrix}
\alpha^{\uparrow}\\ \alpha^{\downarrow}
\end{bmatrix}$, 
$\bfbet=\begin{bmatrix}
\beta^{\uparrow}\\ \beta^{\downarrow} 
\end{bmatrix}$ and $k=\sqrt{2m(E-v_0)/\hbar^2-k_x^2-k_z^2}$.
 
\begin{figure}[tp]
\begin{center}
\includegraphics[width=0.9\linewidth]{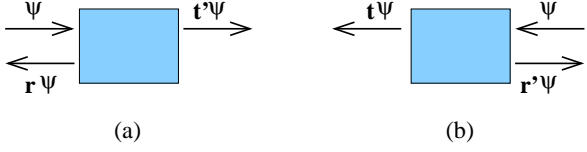}
\end{center}
\caption{\footnotesize Schematic diagram indicating transmitted and reflected wavefunctions for an electron incident from the left (a) and right (b).}
\label{fig1}
\end{figure}

Now consider a general multilayered medium consisting of slabs of homogeneous material stacked in the $y$-direction, with perfect interfaces lying in the $xz$-plane. 
 Let the out-of-plane $k$-vector in the $n$-th layer be $k^{\updownarrow}_n$, and the exchange angle (in the $xz$-plane) be $\theta_n$.
 Furthermore, let $y_{n,n+1}$ be the position of the boundary between the $n$th and $(n+1)$th layer. Then the matching of the wavefunctions, $\bs{\psi}_n$ and $\bs{\psi}_{n+1}$, and their derivatives at this boundary give rise to four simultaneous equations which can be solved to give $\bs{\psi}_n$ in terms of $\bs{\psi}_{n+1}$.
 This procedure can be neatly expressed in terms of transfer matrices $\bfT$ acting on the wavefunction amplitude vector $\bs{\Psi}$. 
 In particular, if we define the transfer matrix ${\bf T}_{n,n+1}$ between neighbouring slabs $n$ and $n+1$ by
 \begin{equation}
 \bs{\Psi}_n={\bf T}_{n,n+1}\bs{\Psi}_{n+1} \quad \text{where} \quad
 \bs{\Psi}_n =\left[
\alpha^{\uparrow}_n,\alpha^{\downarrow}_n,\beta^{\uparrow}_n,\beta^{\downarrow}_n \right]^T
\label{eq:T12} 
\end{equation}  
then it is straightforward to verify that
\begin{equation}
{\bf T}_{n,n+1} = {\bf X}^{-1}(k_n^{\uda},y_{n,n+1}).{\bf S}(\theta_n-\theta_{n+1}).{\bf X}(k_{n+1}^{\uda},y_{n,n+1}),
\label{eq:Tii1}
\end{equation}
 where 
 \begin{equation}
{\bf X}(k^{\uda},y) = 
 \begin{bmatrix}
\bfe  & \bfe^{-1}\\
i \bfk\, \bfe  &  -i \bfk\, \bfe^{-1}
\end{bmatrix}
\quad \text{and} \quad 
 {\bf S}(\theta)=\begin{bmatrix}
 {\bf s}(\theta) & \mathbf{0} \\ \mathbf{0} & {\bf s}(\theta) 
 \end{bmatrix},\label{eq:X}
\end{equation}   
 with 
  $\bfk=\diag[k^{\ua},k^{\da}]$ and $\bfe=\diag[e^{i k^{\ua} y},e^{i k^{\da} y}]$.
  (Here $\diag[a,b]$ indicates a $2 \times 2$ diagonal matrix with elements $a$ and $b$.).
%
%
More generally, the transfer matrix ${\bf T}_{n,m}$  between any two slabs $n$ and $m>n$, is now given in terms of the transfer matrix between neighbouring slabs:
\begin{eqnarray}
&&{\Psi}_n={\bf T}_{n,m}\bs{\Psi}_{m} \cr 
\text{where} \quad &&{\bf T}_{n,m}={\bf T}_{n,n+1}.{\bf T}_{n+1,n+2}\ldots{\bf T}_{m-1,m}.  \nonumber
\end{eqnarray}
There is another more general approach which reveals the structure of the transfer matrices in terms of the reflection and transmission matrices.
This is a generalisation of the approach used in for example ref.\cite{datta1997electronic}, and similar to the one used by Waintal {\em et. al.} \cite{Waintal2000}.
Suppose that we have an electron incident on an arbitrary junction from the left as in Fig~\ref{fig1}(a). If the incoming electron has wavefunction $\bs{\psi}$, then we represent the reflected wavefunction by ${\bf r}\bs{\psi}$ and the transmitted wavefunction by ${\bf t'}\bs{\psi}$.
Likewise if we have an incident electron from the right with wavefunction $\bs{\psi}$, then we represent the reflected wavefunction by ${\bf r'}\bs{\psi}$ and the transmitted wavefunction by ${\bf t}\bs{\psi}$.  Here ${\bf r}$ and ${\bf r'}$ are the left and right $2 \times 2$ reflection matrices, while ${\bf t}$ and ${\bf t'}$ are the left and right $2 \times 2$ transmission matrices respectively.
From these definitions we deduce that the $4 \times 4$ transfer matrix for the system must have the form
\begin{equation}
\mathbf{T} = 
\begin{bmatrix}
\mathbf{t}'^{-1} & -\mathbf{t}'^{-1}\mathbf{r}'  \\ \mathbf{r}\mathbf{t}'^{-1} & \mathbf{t}  -  \mathbf{r} \mathbf{t}'^{-1}\mathbf{r}' 
\end{bmatrix}.\label{eq:Tdef}
\end{equation}
Applying this form to $T_{n,n+1}$ it is straightforward to show that for $\theta_n=\theta_{n+1}=0$, 
\begin{eqnarray}
\left. \bft'_{n,n+1}\right|&=&2\, \diag\hspace{-0.03in} \left[ \tfrac{k_n^{\ua}}{k_n^{\ua}+k_{n+1}^{\ua}} e^{-i(k_{n+1}^{\ua}-k_n^{\ua}) y_{n,n+1}} ,  \ua \lra \da \right]\label{eq:tpii1}\\
\left. \bft_{n,n+1}\right|
&=&2\, \diag\hspace{-0.03in} \left[ \tfrac{k_{n+1}^{\ua}}{k_n^{\ua}+k_{n+1}^{\ua}} e^{-i(k_{n+1}^{\ua}-k_n^{\ua})y_{n,n+1}} ,  \ua \lra \da \right] \label{eq:tii1}\\
\left. \bfr'_{n,n+1}\right|&=&-\diag \left[\hspace{-0.03in} \tfrac{k_n^{\ua}-k_{n+1}^{\ua}}{k_n^{\ua}+k_{n+1}^{\ua}} e^{-2i k_{n+1}^{\ua} y_{n,n+1}} ,  \ua \lra \da \right]\label{eq:rpii1}\\
\left. \bfr_{n,n+1}\right|&=&\diag\hspace{-0.03in} \left[ \tfrac{k_n^{\ua}-k_{n+1}^{\ua}}{k_n^{\ua}+k_{n+1}^{\ua}} e^{2i k_n^{\ua} y_{n,n+1}} ,  \ua \lra \da \right] \ .\label{eq:rii1}
\end{eqnarray}
Here the vertical bar on the left-hand side reminds us that $\theta_n=\theta_{n+1}=0$, and 
 $\diag[a,\ua \lra \da]$ indicates a $2 \times 2$ diagonal matrix whose second diagonal element is the same as the first but with $\ua$ and $\da$ interchanged.
We now employ charge-current conservation to deduce some general algebraic properties of the reflection and transmission matrices. 
In particular, charge-current conservation between any two conducting layers $n$ and $m$ implies:
\[\sum_{\nu=\ua,\da}  k_n^{\nu}(|\bfalp_n^{\nu}|^2-|\bfbet_n^{\nu}|^2) = \sum_{\nu=\ua,\da}  k_m^{\nu}(|\bfalp_m^{\nu}|^2-|\bfbet_m^{\nu}|^2).\]
If we define
\[ \bs{\bfK}_n = \diag[k_n^{\ua},k_n^{\da},-k_n^{\ua},-k_n^{\da} ],\]
then this can be written 
\[\bs{\Psi}_n^{\dagger} \bs{\bfK}_n \bs{\Psi}_n = \bs{\Psi}_m^{\dagger} \bs{\bfK}_m \bs{\Psi}_m, \]
so that
$\bs{\Psi}_m^{\dagger}\bfT_{n,m}^{\dagger} \bs{\bfK}_n \bfT_{n,m}\bs{\Psi}_m= \bs{\Psi}_m^{\dagger} \bs{\bfK}_m \bs{\Psi}_m$
and hence
\begin{equation}
\bfT_{n,m}^{\dagger} \bs{\bfK}_n \bfT_{n,m}=\bs{\bfK}_m.\label{eq:Tinv}
\end{equation}
From this and Eq.(\ref{eq:Tdef}) we deduce that
\begin{eqnarray}
{\bft'}_{n,m}^{\dagger} \bfk_m \bft'_{n,m} + \bfr_{n,m}^{\dagger} \bfk_n \bfr_{n,m}&=&\bfk_n \label{eq:rtid1}\\
\bft_{n,m}^{\dagger} \bfk_n \bft_{n,m} + {\bfr'}_{n,m}^{\dagger} \bfk_m \bfr'_{n,m}&=&\bfk_m \label{eq:rtid2}\\
\bft_{n,m} \bfk_m^{-1} \bft_{n,m}^{\dagger} + {\bfr}_{n,m} \bfk_n^{-1} \bfr_{n,m}^{\dagger}&=&\bfk_n^{-1} \label{eq:rtid2a}\\
{\bft'}_{n,m} \bfk_n^{-1} {\bft'}_{n,m}^{\dagger} + {\bfr'}_{n,m} \bfk_m^{-1} {\bfr'}_{n,m}^{\dagger}&=&\bfk_m^{-1} \label{eq:rtid2b}
\end{eqnarray}
where $\bfk_n=\diag[k_n^{\ua},k_n^{\da}]$.

This is as far as we can go for general reflection and transmission matrices.
However, in this communication we will be interested in magnetic multilayers, composed of non-magnetic (NM) and ferromagnetic (FM) layers, sandwiched consecutively so that each FM layer has NM layers either side of it. We also assume that the exchange field is rotated in the $xz$-plane.
For such systems the reflection and transmission matrices have additional symmetry properties. In particular, in the Appendix we show that (see Eq's (\ref{eq:rtrans}), (\ref{eq:ttrans}))
\begin{eqnarray}
{\bfr'}^T_{n,m} &=&\bfr'_{n,m}\quad ,\quad\bfr_{n,m}^T =\bfr_{n,m}\\
{\bft'}^T_{n,m} &=&  k_n k_m^{-1 }\bft_{nm}.
\end{eqnarray}
when $n$ and $m$ are non-magnetic layers.


Note that when dealing with charge currents, it is more usual to define the transfer matrix in terms of the charge-current-amplitude,
 $\tilde{\bs{\Psi}}$, rather than the wavefunction-amplitude:
\begin{eqnarray}
\tilde{\bs{\Psi}}_n&=&\tilde{{\bf T}}_{n,m}\tilde{\bs{\Psi}}_{m} \quad \text{where} \quad \tilde{\bs{\Psi}}_n = \bs{\Xi}_n\bs{\Psi}_n \cr
\bs{\Xi}_n&=&\diag \left[ \sqrt{k_n^{\ua}},\sqrt{k_n^{\da}},\sqrt{k_n^{\ua}},\sqrt{k_n^{\da}} \right]. \nonumber
\end{eqnarray}
Our transfer matrix is therefore related to the usual one by $\tilde{{\bf T}}_{n,m}=\bs{\Xi}_n\bfT_{n,m}\bs{\Xi}_m^{-1}$, and our reflection and transmission matrices are likewise a rescaling of the usual ones $\tilde{\bfr}, \tilde{\bft},\tilde{\bfr}',\tilde{\bft}'^2$. In terms of these, providing everything commutes, Eq's (\ref{eq:rtid1})--(\ref{eq:rtid2b}) take the more usual form $|\tilde{\bfr}|^2+|\tilde{\bft}|^2=1$, $|\tilde{\bfr}'|^2+|\tilde{\bft}'|^2=1$, $|\tilde{\bfr}|^2=|\tilde{\bfr}'|^2$ and $|\tilde{\bft}|^2=|\tilde{\bft}'|^2$. 
Despite the simplicity of these particular equations, because we need to deal with both conductors (real k) and insulators (imaginary k), in what follows much of the algebra is more straightforward if we work with the `rescaled' reflection and transmission matrices $\bfr,\bfr',\bft,\bft'$ introduced here.
The end results are, of course, the same no matter which definitions are used.

Having set up the framework for calculating the wavefunction in an arbitrary slab of a multilayer, let us now move on to the calculation of the spin current in the standard multilayer switching geometry.

We note that throughout this communication we use the following convention when we are able:   bold upper case Latin characters refer to $4 \times 4$ matrices;  bold lower case Latin characters refer to $2 \times 2$ matrices; bold upper case Greek characters refer to $4$ dimensional vectors; bold lower case Greek characters refer to $2$ dimensional vectors. Of course there are one or two exceptions like $\bfsig$ which by convention denotes the $2 \times 2$ Pauli matrices.

\section{The Switching Geometry and Landauer Formalism}\label{sec:land}
\begin{figure}[tp]
\begin{center}
\includegraphics[width=0.9\linewidth]{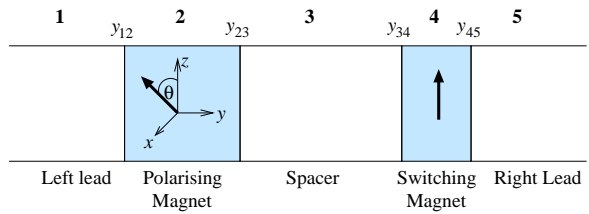}
\end{center}
\caption{\footnotesize The standard multilayer switching geometry.}
\label{fig2}
\end{figure}

The standard multilayer geometry for current induced switching is shown in Fig.~\ref{fig2}.
It consists of a five slab multilayer consisting of two semi-infinite non-magnetic leads and a non magnetic spacer, which separates a polarizing magnet (PM) and a switching magnet (SM). The magnetization of the PM is rotated by an angle $\theta$ in the $xy$-plane, while the magnetization of the SM is parallel to the $z$-axis.  
We label each of the consecutive slabs $1,2,3,4,5$, and assume
that the entire structure is homogeneous in the $xy$-plane, with perfect interfaces at positions $y_{n,n+1}$, where $i=1,2,3,4$.

Our ultimate aim is to calculate the spin current in the spacer: slab 3. This is defined locally in terms of the spacer wavefunction $\bs{\psi}$ by
\[
j_i = \frac{\hbar^2}{4mi}\left(\bs{\psi}^{\dagger}\bs{\sigma}_i \bs{\psi}' -  \bs{\psi}'^{\dagger}\bs{\sigma}_i \bs{\psi}  \right).
\]
Note that if we define $\bs{\sigma}_0=(2e/\hbar)\,\mathbf{1}$, then $j_0$ is simply the charge current.
Within a non-magnetic slab, all components of the spin current are conserved, and using equation~(\ref{eq:wf1}), in the case where non-magnetic slab is a conductor ($k \in \mathbb{R}$), the spin current becomes
\begin{equation}
j_i = k(\bs{\alpha}^{\dagger}\bs{\sigma}_i\bs{\alpha} - \bs{\beta}^{\dagger}\bs{\sigma}_i\bs{\beta}).\label{eq:sc1}
\end{equation}
We will calculate the spin current within the Landauer formalism \cite{Waintal2000,Edwards2006273}. 
In particular we assume that the left ($L$) and right ($R$) leads are connected to two non-magnetic reservoirs whose electron distributions are characterised by Fermi functions $f_L=f(\varepsilon-\mu_L)$ and $f_R=f(\varepsilon-\mu_R)$, and whose chemical potentials $\mu_L$ and $\mu_R$ are displaced by an infinitesimal bias, so that $\mu_L-\mu_R=eV_b$. 
Because the bias is infinitesimal the one-electron states can be calculated from the Schr\"{o}dinger equation neglecting $V_b$.
We take the global spin quantization axis of our system to be the $z$-direction, and label the up and down spin-eigenvectors by $\ua$ and $\da$ respectively. 
Following the Landauer methodology, we assume that
electrons of both spin orientations are emitted from the reservoirs at all momenta $(k_x,k_z)$ and energies $\varepsilon$ up to their Fermi level, so that $i$th component of the total spin current at any point in the system is given by 
\begin{equation}
J_i^{\text{tot}} = \sum_{k_x,k_z} \int d \varepsilon  \sum_{\nu=\ua,\da}  f_L \orj_i^{\nu}(k_x,k_z,\varepsilon) +  f_R \olj_i^{\nu}(k_x,k_z,\varepsilon).\nonumber
\end{equation}
Here $\orj_i^{\nu}$ and $\olj_i^{\nu}$ are the individual spin currents arising from electrons of spin $\nu$ incident from the left and right reservoirs respectively.
Defining 
\[\olrj_i=\olrj_i^{\ua} + \olrj_i^{\da}\]
and symmetrising over the Fermi functions, this can be written
\[
J_i^{\text{tot}} = \sum_{k_x,k_z} \int d \varepsilon   (f_L+f_R) [\orj_i+\olj_i] +  (f_L-f_R)[\orj_i-\olj_i].
\]
In the absence of bias only the first term (in $\olrj_y$) remains, so it is identified as the ``static''-spin current, responsible for the oscillatory exchange coupling between two ferromagnets.
In fact, later (Eq.~\ref{eq:lrscequiv}) we show explicitly that, if $i=0,x,z$ then $\orj_i = -\orj_i$, so only the $y$-component of spin current contributes to the oscillatory exchange coupling.

  For infinitesimal bias $\mu_{\substack{L \\ R}}=\mu \pm \tfrac12 eV_b$, then $f_L-f_R \sim -eV_b\, \delta(\varepsilon-\mu)$, so the second term is proportional to $V_b$ and so we refer to this the ``transport-spin current'':
\[
J_i^{\text{tr}} = e V_b \sum_{k_x,k_z} j_i^{\text{tr}}(k_x,k_z)
\ \ \text{where}\ \ 
j_i^{\text{tr}} = \sum_{\nu=\ua,\da} \orj_i^{\nu} -  \olj_i^{\nu} .
\]
In the limit of large spacer thickness this term will eventually dominate since it remains finite whilst the exchange coupling decays to zero.

In general all four components of this spin current are non-zero. When $i=0$ it is the charge current, when $i=x$ this is referred to as the in-plane spin current, and when $i=y$ it is referred to as the out-of-plane spin current. 
In what follows we shall explore the origin and symmetry properties of the spin current, particularly the out-of-plane component which was for some time misunderstood. 
In particular, we shall use the Landauer formulation prove that it vanishes when there is exact matching of potentials in one spin band, or when there is reflection symmetry in the system.
The former property explains why Slonczewski's original calculation \cite{Slonczewski1996} deduced that there was no out-of-plane spin transfer to the switching magnet.
The latter property has been verified previously by numerical calculations on realistic systems \cite{Edwards2005}, \cite{Butler2010}, and explains why the authors or Ref. \cite{Butler97} originally concluded that the out-of-plane transport-spin current $J_y^{\text{tr}}$ was a quadratic function of the applied bias.

\section{Origin of out-of-plane spin current}\label{sec:origin}
In this section we illuminate the origin of the in-plane and out-of-plane spin current in the spacer.
We do this in several stages, by splitting the system into two parts: Lead/PM/Spacer and Spacer/SM/Lead.

Similar studies have been performed previously in special cases \cite{Stiles2002,Ralph2008}, however some of the results derived here are required for the remainder of the paper.

\subsection{Lead / PM / Spacer}
First we examine what happens when electrons of spin $\nu=\ua$ and $\nu=\da$ are injected into the left lead and emerge from the PM into the spacer.
Note that at this point we are completely ignoring the influence of the switching magnet and the right lead.  Later we will include their effect.

To calculate the wavefunction in the spacer, we need to determine the transfer matrix $\bfT_{13}$, between the lead (slab 1) and the spacer (slab 3).  From Eq.~(\ref{eq:Tii1}) we have
\begin{eqnarray}
\bfT_{13}(\theta)&=&\bfT_{12}.\bfT_{23}=
{\bf S}^{-1}(\theta).\bfT_{13}(0).{\bf S}(\theta),\nonumber
\end{eqnarray}
where we have used the fact that slab 1 and 3 are non-magnetic.
Furthermore since the reflection and transmission matrices are simply related to the $2 \times 2$ submatrices of $\bfT$, it follows that
\begin{equation}
\bft_{13}'(\theta)={\bf s}^{-1}(\theta).\bft'_{13}(0).{\bf s}(\theta) \quad \text{similarly for}\quad \bft, \bfr, \bfr'. \label{eq:thdep}
\end{equation}
This shows that the $\theta$ dependence of the Lead/PM/Spacer sub-system is very simple, and for the most part we can work with transmission and reflection matrices at $\theta=0$, and put the $\theta$ dependence back in at the end.

For simplicity, we now take the position of the first and second interfaces to be $y_{12}=0$ and $y_{23}=L$, and take the wavevector in the spacer and lead to be identical $k_1=k_3=k$. We also write the wavevector in the PM as $k_2^{\ua}=k^{\ua}$ and $k_2^{\da}=k^{\da}$.
Then the wavefunction in the spacer (slab 3) is $\overrightarrow{\bs{\psi}}_3=\overrightarrow{\bs{\alpha}}_3\, e^{iky}$ and so the spin current is
\begin{equation}
\overrightarrow{j}_i=k \overrightarrow{\bs{\alpha}}_3^{\dagger}\bs{\sigma}_i\overrightarrow{\bs{\alpha}}_3,
\label{eq:j31}
\end{equation}
\begin{figure}[tp]
\begin{center}
\includegraphics[width=0.9\linewidth]{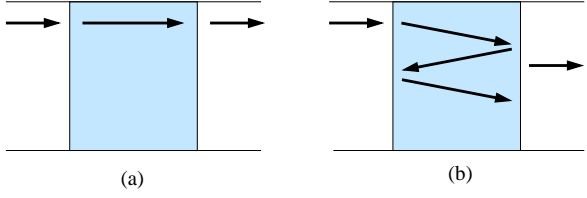}
\end{center}
\caption{\footnotesize Schematic depicting transmission through the PM, (a) without reflections (b) including all reflections.}
\label{fig2ab}
\end{figure}%
where $\overrightarrow{j}_i$ indicates that we are only considering the current contributed by electrons incident from the left lead. Unless otherwise stated, we shall only be dealing with the current in the spacer (layer 3), therefore the layer index will not be used explicitly in the notation for current components. Let us first examine the lowest order contribution to this spin current i.e. the component of the wavefunction which passes through the PM into the spacer without any reflections as depicted in Fig.\ref{fig2ab}(a). Then
\begin{equation}
\overrightarrow{\bfalp}_3=\bfs^{-1}(\theta)\, \bft'_{12}(0)\, \bft'_{23}(0)\, \bfs(\theta)\, \overrightarrow{\bfalp}_1,
\label{eq:alp31}
\end{equation}
Combining Equations (\ref{eq:tpii1}), (\ref{eq:j31}) and (\ref{eq:alp31}) gives the transport spin current in the spacer, originating from an electron of spin $\overrightarrow{\bfalp_1}$ coming from the left lead, ignoring any reflections in the PM.  
In particular for an incident up-spin electrons $\overrightarrow{\bfalp}_1=\overrightarrow{\bfalp}_1^{\ua}=\begin{bmatrix}
1\\0\end{bmatrix}$  we obtain
\begin{eqnarray}
\overrightarrow{j}_x^{\ua}&=&16 k^3 \sin \theta\left[\frac{{k^{\ua}}^2\cos^2 \tfrac{\theta}{2}}{(k+k^{\ua})^4}  -   \frac{{k^{\da}}^2\sin^2 \tfrac{\theta}{2}}{(k+k^{\da})^4}  \right] \qquad \cr
 &&\qquad-\ \frac{8 k^3 k^{\ua}k^{\da} \sin 2\theta  \cos L(k^{\ua}-k^{\da}) }{(k+k^{\ua})^2 (k+k^{\da})^2}\cr
\overrightarrow{j}_y^{\ua}&=&\frac{16 k^3 k^{\ua}k^{\da} \sin \theta  \sin L(k^{\ua}-k^{\da}) }{(k+k^{\ua})^2 (k+k^{\da})^2}\nonumber
\end{eqnarray}
While for incident down-spin electrons  $\bfalp_1=\bfalp_1^{\da}=\begin{bmatrix}
0\\1\end{bmatrix}$ we get
\begin{eqnarray}
\overrightarrow{j}_x^{\da}&=&16 k^3 \sin \theta\left[\frac{{k^{\ua}}^2\sin^2 \tfrac{\theta}{2}}{(k+k^{\ua})^4}  -   \frac{{k^{\da}}^2\cos^2 \tfrac{\theta}{2}}{(k+k^{\da})^4}  \right] \qquad \cr
&&\qquad+\ \frac{8 k^3 k^{\ua}k^{\da} \sin 2\theta  \cos L(k^{\ua}-k^{\da}) }{(k+k^{\ua})^2 (k+k^{\da})^2}
\label{eq:jxd0}\cr
\overrightarrow{j}_y^{\da}&=&-\overrightarrow{j}_y^{\ua}.\nonumber
\end{eqnarray}
Notice that $\overrightarrow{j}_x^{\ua}$ and $\overrightarrow{j}_x^{\da}$ consists of two distinct terms, one which is independent of the PM thickness $L$, and one which depends sinusoidally on $L$. 
These have a clear physical interpretation.
The term independent of $L$ is due to wavefunction matching of the incident electrons at the interfaces. It vanishes when $\theta=0$ or $\pi$ i.e. when the moment of the PM is parallel or anti-parallel to the incident electrons. 
The term which depends sinusoidally on $L$ is due to spin precession of the electron as is passes through the PM. We note that it vanishes when $\theta=0$ or $\pi$ corresponding to no precession, or when $\theta=\pi/2$ corresponding to pure out-of-plane precession.

On the other hand, $j_y^{\ua}$ and $j_y^{\da}$ consist only of a precessional term, because the PM is rotated only in-plane. This term vanishes only when $\theta=0$ or $\pi$  corresponding to no precession and has a maximum at $\theta=\pi/2$  corresponding to pure out-of-plane precession.

In both cases up-spin electrons precess in an equal but opposite direction to down-spin electrons, so the total spin current in the spacer is
\begin{eqnarray}
\overrightarrow{j}_x&=& \overrightarrow{j}_x^{\ua} + \overrightarrow{j}_x^{\da}\cr
&=& 16 k^3 \sin \theta\left[\frac{{k^{\ua}}^2}{(k+k^{\ua})^4}  -   \frac{{k^{\da}}^2}{(k+k^{\da})^4}  \right] \cr
\overrightarrow{j}_y&=& \overrightarrow{j}_y^{\ua} + \overrightarrow{j}_y^{\da} = 0 \ .\nonumber
\end{eqnarray}
So we see that the in-plane spin current arises from wavefunction matching between the lead and PM, which is rotated in-plane. 
There is no out-of-plane spin current because the PM is not rotated out-of-plane.

When the effect of reflections in the PM are included,  as depicted in Fig.\ref{fig2ab}(b), the explanation given above remains qualitatively the same. In particular, $j_x^{\ua}$ and $j_x^{\da}$ can be shown to be composed of a wavefunction matching term which is independent of the thickness $L$, plus a precessional term which is quasi-periodic in $L$ with angular frequencies $2Lk^{\ua}$, $2Lk^{\da}$, $L(k^{\ua}-k^{\da})$, $L(k^{\ua}+k^{\da})$ and higher harmonics. 
Also it can be shown that $\overrightarrow{j}_y^{\ua}=-\overrightarrow{j}_y^{\da}$ is a purely precessional term which is a quasi-periodic function of $L$.

Despite the fact that individual spin currents have rather complicated expressions when all reflections in the PM are included, the sum $\overrightarrow{j}_i^{\ua} + \overrightarrow{j}_i^{\da}$ is rather simple. 
For this case Equation~(\ref{eq:alp31}) is replaced by
 \begin{equation}
\overrightarrow{\bfalp}_3=\bfs^{-1}(\theta)\, \bft'_{13}(0) \bfs(\theta)\, \overrightarrow{\bfalp}_1,
\label{eq:alp3}
\end{equation}
where $\bft'_{13}(0)$ is the transmission matrix obtained from the transfer matrix $\bfT_{13}$ evaluated at $\theta=0$.
So from Eq.~(\ref{eq:j31})
\begin{eqnarray}
\overrightarrow{j}_i&=& \overrightarrow{j}_i^{\ua} + \overrightarrow{j}_i^{\da} \cr
&=& k \sum_{\nu=\ua,\da} \overrightarrow{\bfalp}_1^{\nu\, \dagger}\, \bfs(-\theta)\, \bft'_{13}(0)^{\dagger} \,  \bfs(\theta)
\bs{\sigma}_i\bfs^{-1}(\theta)\, \bft'_{13}(0) \bfs(\theta)\, \overrightarrow{\bfalp}_1^{\nu}\cr
&=& k\  \tr \left(\bft'_{13}(0) \bft'_{13}(0)^{\dagger} \,  \bfs(\theta) \bs{\sigma}_i\bfs^{-1}(\theta)\,  \right).\label{eq:j3}
\end{eqnarray}
This form clearly demonstrates that the total spin current in the spacer is independent of the direction of polarization of incident electrons: since if $\overrightarrow{\bfalp}_1^{\nu}(\omega)=\bfs(\omega)\overrightarrow{\bfalp}_1^{\nu}$, then
$\sum_{\nu} \overrightarrow{\bfalp}_1^{\nu}(\omega)^{\dagger}\mathbf{m}\overrightarrow{\bfalp}_1^{\nu}(\omega) = \sum_{\nu} \overrightarrow{\bfalp}_1^{\nu\, \dagger}\mathbf{m}\overrightarrow{\bfalp}_1^{\nu} = \tr(\mathbf{m})$ for any matrix $\mathbf{m}$.

Notice that we can rewrite Eq.~(\ref{eq:j3}) in the form $\overrightarrow{j}_i=k \sum_{\nu} \overrightarrow{\bfgam}^{\nu\, \dagger} \bs{\sigma}_i \overrightarrow{\bfgam}^{\nu}$ where $\overrightarrow{\bfgam}^{\nu}=\bfs^{-1}(\theta) \bft'_{13}(0) \overrightarrow{\bfalp}_1^{\nu} $. 
Since $\bft'_{13}(0)$ is diagonal then $\overrightarrow{\bfgam}^{\ua}$ corresponds to an electron polarized in the  in-plane $-\theta$ direction, multiplied by a weight  $[\bft'_{13}]_{11}$. 
Whereas $\overrightarrow{\bfgam}^{\da}$ corresponds to an electron polarized in the  in-plane $\theta$ direction multiplied by a weight  $[\bft'_{13}]_{22}$. 
Either using this observation, or by explicit calculation from Eq.~(\ref{eq:j3}) we immediately obtain
\begin{eqnarray}
\overrightarrow{j}_x&=& k \sin \theta  \left(\left[\bft'_{13}(0) \bft'_{13}(0)^{\dagger}\right]_{11} - \left[\bft'_{13}(0) \bft'_{13}(0)^{\dagger}\right]_{22} \right)\cr
\overrightarrow{j}_y&=& 0.
\label{eq:j3xy}
\end{eqnarray}

So, the effect of the PM is to rotate the incident up and down ($z$-polarized) spin electrons by an angle $\theta$ in-plane and to give each a different weight, depending on the matching of wavefunctions across the PM.
In addition there is a precessional term, but this is equal and opposite for up and down spin electrons.
The total spin current of electrons emerging from the PM therefore only has an in-plane component due to wavefunction matching.

Let us now move on to consider the effect of the switching magnet.

\subsection{Spacer / SM / Lead}
\begin{figure}[tp]
\begin{center}
\includegraphics[width=0.5\linewidth]{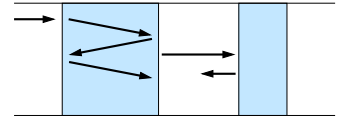}
\end{center}
\caption{\footnotesize Schematic depicting reflection off the $\text{spacer}\vert\text{SM}$  interface only.}
\label{fig3}
\end{figure}

We begin by calculating the total spin current in the spacer, for electrons incident from the left hand lead, after passing through the PM and then reflecting off the $\text{spacer}\vert\text{SM}$ ($3\vert 4$) interface only, as depicted in Fig.~\ref{fig3}. This is in effect a 1st order reflection from the SM, we will examine higher orders shortly.
Then the total spin current in the spacer is
\begin{equation}
\overrightarrow{j}_i = k \sum_{\nu=\ua,\da} (\overrightarrow{\bfalp}_3^{\nu\,\dagger}\bfsig_i\overrightarrow{\bfalp}_3^{\nu} - \overrightarrow{\bfbet}_3^{\nu\,\dagger}\bfsig_i\overrightarrow{\bfbet}_3^{\nu}),
\label{eq:j3new}
\end{equation}
where $\overrightarrow{\bfalp}_3^{\nu}$ is given by Eq.(\ref{eq:alp3}) and $\overrightarrow{\bfbet}_3^{\nu}=\bfr_{34}\overrightarrow{\bfalp}_3^{\nu}$.
The first term, due to electrons which have passed through the PM into the spacer, has already been calculated and analysed in Eq's.(\ref{eq:j3}) and (\ref{eq:j3xy}).
The second term, which we call $\overrightarrow{j}_i^{\,\text{R1}}$, is due to electrons reflected from the $3\vert 4$ interface.
\begin{eqnarray}
\overrightarrow{j}_i^{\,\text{R1}}&=&-k \sum_{\nu=\ua,\da}   \overrightarrow{\bfbet}_3^{\nu\,\dagger}\bfsig_i\overrightarrow{\bfbet}_3^{\nu}\cr
&=&-k \sum_{\nu=\ua,\da}   \overrightarrow{\bfalp}_3^{\nu\, \dagger} \bfr_{34}^{\dagger}\, \bfsig_i \, \bfr_{34}\, \overrightarrow{\bfalp}_3^{\nu}\cr
&=& - k\, \tr ( \mathbf{m} \bs{\sigma}_i ), \label{eq:jR1}
\end{eqnarray}
where
\[ 
\mathbf{m}=\bfr_{34}\bfs^{-1}(\theta) \bft'_{13}(0) {\bft'}_{13}^{\dagger}(0)  \bfs(\theta) \bfr_{34}^{\dagger}.\]
We can derive several conclusions from this result. 
Firstly, the reflected in-plane spin current 
\[\overrightarrow{j}_x^{\,\text{R1}} \ne 0 \quad \text{ in general.}\]
Further, we note that from Eq.(\ref{eq:rii1}), 
$\bfr_{34}=e^{2i k_3 y_{3,4}}\tilde{\bfr}_{34}$, where $\tilde{\bfr}_{34}=
\diag \left[ \tfrac{k_3-k_{4}^{\ua}}{k_3+k_{4}^{\ua}} \ ,\  \ua \lra \da \right]$. 
Hence the matrix $\mathbf{m}=\tilde{\bfr}_{34}\bfs^{-1}(\theta) \bft'_{13}(0) \bft'_{13}(0)^{\dagger}  \bfs(\theta) \tilde{\bfr}_{34}^{\dagger}$, and since $\bft'_{13}(0)$ is diagonal, then $\mathbf{m}$ is real if $\tilde{\bfr}_{34}$ is. 
Since $\bfsig_y$ is pure imaginary and we are assuming that the spacer is not an insulator, then we deduce that
\begin{eqnarray}
 \overrightarrow{j}_y^{\,\text{R1}} &=& 0 \quad \text{if $k_4^{\ua}$ and $k_4^{\da}$ are real.}\cr
 \overrightarrow{j}_y^{\,\text{R1}} &\ne& 0 \quad \text{if $k_4^{\ua}$ or $k_4^{\da}$ is pure imaginary.}\nonumber
\end{eqnarray}
So after reflection from the $\text{spacer}\vert\text{SM}$ interface, an out of plane spin current only arises if the SM is a half-metallic ferromagnet or a magnetic insulator.

\begin{figure}[tp]
\begin{center}
\includegraphics[width=0.5\linewidth]{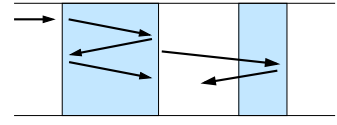}
\end{center}
\caption{\footnotesize Schematic depicting reflection off the $\text{SM}\vert\text{lead}$  interface only.}
\label{fig4}
\end{figure}

For the case where the SM is a metallic ferromagnet, we need to consider the spin current in the spacer after a 2nd order reflection from the SM i.e.  for electrons incident from the left hand lead, which pass through the PM, then are reflected once off the $4\vert 5$ interface back into the spacer, as depicted in Fig.~\ref{fig4}. 
The reflection matrix for this process is given by
\begin{eqnarray}
 {\bfr}_{35}^\text{R2}&=&\bft_{34}\bfr_{45}\bft'_{34}\cr
 &=&4 k_3 e^{2i k_3 y_{3,4}}\diag \left[k_4^{\ua} e^{2i k_4^{\ua} (y_{4,5}-y_{3,4})} \tfrac{k_4^{\ua}-k_{3}}{(k_4^{\ua}+k_{3})^3} \ ,\  \ua \lra \da \right] \nonumber
\end{eqnarray} 
where we have used Eq's.(\ref{eq:tpii1})-(\ref{eq:rii1}).
The spin current arising from this second order reflection, $\overrightarrow{j}_i^{\,\text{R2}}$, has exactly the same form as that from the first order reflection $\overrightarrow{j}_i^{\,\text{R1}}$ given in Eq.(\ref{eq:jR1}), but with $\bfr_{34}$ replaced by $\bfr_{35}^\text{R2}$.
However now, the matrix $\mathbf{m}$ cannot be real because of the presence of the $e^{2i k_4^{\uda} (y_{4,5}-y_{3,4})}$ term in  $\bfr_{35}^\text{R2}$.
Hence both $\overrightarrow{j}_x^{\,\text{R2}}$ and $\overrightarrow{j}_y^{\,\text{R2}}$ are non-zero.

The interpretation of the above results is that (for electrons incident from the left-hand lead) the total spin current in the spacer after they emerge from the PM only has an in-plane component.
The out-of-plane component arises because this in-plane component precesses in the SM and is reflected back.
In the case where the SM is a conductor, in order for precession to take place, this reflection has to occur off the $4\vert 5$ interface. In the case where the SM is a half-metallic ferromagnet or a magnetic insulator, $\bfr_{34}$ includes reflection which occurs throughout the SM because of the decay of the wavefunction.

In fact, it is straightforward to show that $\overrightarrow{j}_y^{\,\text{R1}}$ and $\overrightarrow{j}_y^{\,\text{R2}}$ have the form
 \begin{eqnarray}
  \overrightarrow{j}_y^{\,\text{R1}}&=&k\sin(\theta)(d_1-d_2) \Im (f^{\ua}f^{\da\, *}) \cr
 \overrightarrow{j}_y^{\,\text{R2}}&=&k\sin(\theta)(d_1-d_2)c^{\ua}c^{\da} \sin(L'(k_4^{\ua}-k_4^{\da})) 
  \nonumber
 \end{eqnarray}
 where $L'$ is the SM thickness, $d_i= |[\bft'_{13}(0)]_{ii}|^2$, $f^{\uda}=(k-k_4^{\uda})/(k+k_4^{\uda})$ and $c^{\uda}=-4kk_4^{\uda}(k-k_4^{\uda})/(k+k_4^{\uda})^3$.
So we see that for a metallic SM, $\overrightarrow{j}_y^{\,\text{R2}}$ oscillates sinusoidally about zero as a function of the SM thickness, and so arises from pure precession in the SM.
For half-metallic or insulating SM, the $\overrightarrow{j}_y^{\,\text{R1}}$ does not oscillate as a function of $L'$ although it will oscillate about zero as a function of the spacer thickness.
Furthermore, $\overrightarrow{j}_y^{\,\text{R1}} \rightarrow 0$ as the height of the SM barrier increases, indicating that electrons entering the SM have less time to precess before they are reflected.

Having discussed the origin of the out-of-plane spin current, let us now move to calculating some symmetry properties of the spin current in the spacer.
\section{Symmetry properties of the spin current}\label{sec:symmetry}
In the previous section we were only concerned with the spin current due to electrons incident from the left-hand lead and with low numbers of reflections. 
In what follows, we will need to employ a full calculation and consider the spin current due to electrons incident from both leads and sum over all reflections. 

If we have incident electrons in the left-hand lead, with amplitude $\orbalp_1$, then (summing over all reflections) the right- and left-moving amplitudes $\orbalp_3$  and $\orbbet_3$ in the spacer are given by
\begin{eqnarray}
\orbalp_3&=&\left[ \bft'_{13} +  \bfr'_{13}\bfr_{35}\bft'_{13} +  \bfr'_{13}\bfr_{35}\bfr'_{13}\bfr_{35}\bft'_{13} + \ldots  \right ] \orbalp_1 \nonumber\\
&=&\orbA_3\orbalp_1\cr
\orbbet_3&=&\left[\bfr_{35}\bft'_{13} +  \bfr_{35}\bfr'_{13}\bfr_{35}\bft'_{13} + \ldots  \right ] \orbalp_1 \nonumber\\
&=&\orbB_3\orbalp_1  \nonumber
\end{eqnarray}
where
\begin{eqnarray}
\orbA_3&=\left[ \mathbf{1} -\bfr'_{13} \bfr_{35} \right]^{-1} \bft'_{13}&      \quad ,\quad  \orbB_3=\bfr_{35}\orbA_3.\label{eq:ABr}
\end{eqnarray}
The total spin current in the spacer due to electrons incident from the left is given by Eq.(\ref{eq:j3new}), which can now be written
\begin{eqnarray}
\orj_i&=&k_3\  \tr \left[\orbA_3^{\dagger} \bfsig_i \orbA_3 -  \orbB_3^{\dagger} \bfsig_i \orbB_3\right].\label{eq:jl}
\end{eqnarray}
Likewise, if we have electrons incident from the right-hand lead (slab 5), with amplitude $\olbbet_5$, then the right- and left-moving amplitudes $\olbalp_3$  and $\olbbet_3$ in the spacer are given by $\olbalp_3=\olbA_3\olbbet_5$ and $\olbbet_3=\olbB_3\olbbet_5$ where
\begin{eqnarray}
\olbB_3=\left[\mathbf{1} - \bfr_{35}\bfr'_{13} \right]^{-1} \bft_{35}\quad , \quad
\olbA_3=\bfr'_{13}\olbB_3.\label{eq:ABl}
\end{eqnarray}
The total spin current in the spacer due to electrons incident from the right is 
\begin{eqnarray}
\olj_i&=&k_3\  \tr \left[\olbA_3^{\dagger} \bfsig_i \olbA_3 -  \olbB_3^{\dagger} \bfsig_i \olbB_3\right].\label{eq:jr}
\end{eqnarray}
Eq.'s (\ref{eq:jl}) and (\ref{eq:jr}) explicitly show that the spin currents $\orj_i$ and$\olj_i$ are independent of the polarization axis of the incoming electrons, since  
\[
\olrj_i=k_3\  \sum_{\nu=\ua,\da} \left\langle \nu \right| \olrbA_3^{\dagger} \bfsig_i \olrbA_3 -  \olrbB_3^{\dagger} \bfsig_i \olrbB_3 \left| \nu \right\rangle
\]
and these expressions remain unchanged under $\left| \nu \right\rangle \rightarrow \bfs(\theta)\left| \nu \right\rangle$. 

Furthermore, we note that under an in-plane rotation of the entire system by an angle theta,  $\olrbA_3 \rightarrow \bfs(\theta)^{-1} \olrbA_3\, \bfs(\theta)$, the out-of-plane components $\olrj_y$ remain unchanged, whereas the in-plane components $\olrj_x$ and $\olrj_z$ rotate as expected.
We now explore some symmetry properties of these spin currents.

\subsection{Exact Matching}
First we examine what happens when we have exact matching of potentials, in one spin band, across the system.
This case is sometimes used in model calculations of magnetoresistance and spin current effects (see for example Ref. \cite{Slonczewski1996}) or Ref. \cite{Castro96}) and so deserves attention, particularly because in this special case we will demonstrate that the out of plane spin current vanishes.

Suppose that we have exact matching in, for example, the up-spin band so that $k_1=k_2^{\ua}=k_3=k_4^{\ua}=k_5=k$.  
Then from Eq's.(\ref{eq:rpii1}) and (\ref{eq:rii1}), we see that at $\theta=0$, both $\bfr_{n,n+1}$ and $\bfr'_{n,n+1}$ have the form of a down-spin projection matrix $\bfp^{\da}$
\[
\bfr_{n,n+1}(0) \sim \bfr'_{n,n+1}(0) \propto \bfp^{\da} = \begin{bmatrix}
0 & 0 \cr 0 & 1
\end{bmatrix}.
\]
Furthermore, since 
\begin{eqnarray}
\bfr'_{13}&=&\bfr'_{23}+\bft'_{23}\bfr'_{12}\bft_{23}+
\bft'_{23}\bfr'_{12}\bfr_{23}\bfr'_{12}\bft_{23}+\ldots \nonumber\\
&=&\bfr'_{23}+\bft'_{23}[1-\bfr'_{12}\bfr_{23}]^{-1}\bfr'_{12}\bft_{23} \nonumber
\end{eqnarray}
then this is also proportional to $p^{\da}$ at $\theta=0$.
Similarly for $\bfr_{35}$ and the other $(1,3)$ and $(3,5)$ reflection matrices.
All this is to be expected, since at  $\theta=0$ exact matching implies one component of the spin will `see' a system without interfaces i.e. will have no reflections.

For $\theta \ne 0$, $\bfr'_{13}(\theta)=\bfs^{-1}\,\bfr'_{13}(0)\,\bfs$, $\bft'_{13}(\theta)=\bfs^{-1}\,\bft'_{13}(0)\,\bfs$, and similarly for $\bfr_{13}$ and $\bft_{13}$.
So all the $(1,3)$ reflection and transmission matrices and their hermitian conjugates commute. Hence from Eq.'s  (\ref{eq:rtid1})--(\ref{eq:rtid2b}), we deduce that
\begin{eqnarray}
\bfr_{13}^{\dagger}\bfr_{13}+\bft_{13}^{\dagger}\bft_{13}=\mathbf{1}\label{eq:rtid6}\\
{\bfr'}_{13}^{\dagger}{\bfr'}_{13}+{\bft'}_{13}^{\dagger}{\bft'}_{13}=\mathbf{1} \label{eq:rtid7}\\
\bfr_{13}^{\dagger}\bfr_{13}={\bfr'}_{13}^{\dagger}{\bfr'}_{13} \quad,\quad \bft_{13}^{\dagger}\bft_{13}={\bft'}_{13}^{\dagger}{\bft'}_{13}. \label{eq:rtid8}
\end{eqnarray}
Similar results are true for the $(3,5)$ reflection and transmission matrices.

Now, 
\begin{equation}
\tr [\orbA_3^{\dagger} \bfsig_i \orbA_3 ] = \tr \left[[ \mathbf{1} - \bfr_{35}^{\dagger} {\bfr'}_{13}^{\dagger}]^{-1} \bfsig_i  [\mathbf{1} - \bfr'_{13}\bfr_{35} ]^{-1} \bft'_{13} {\bft'}_{13}^{\dagger}  \right],\label{eq:tmpa1}
\end{equation}
so that using commutativity of $\bft'_{13}$ and ${\bft'}_{13}^{\dagger}$, and Eq.(\ref{eq:rtid7}) we deduce
\begin{eqnarray}
\tr [\orbA_3^{\dagger} \bfsig_i \orbA_3 ] = \tr \left[[ \mathbf{1} - \bfr_{35}^{\dagger} {\bfr'}_{13}^{\dagger}]^{-1} \bfsig_i  [ \mathbf{1}  - \bfr'_{13}\bfr_{35} ]^{-1}\right] \cr -
\tr \left[[ \mathbf{1} -  {\bfr'}_{13}^{\dagger} \bfr_{35}^{\dagger}]^{-1} ({\bfr'}_{13}^{\dagger}\bfsig_i{\bfr'}_{13})  [\mathbf{1} - \bfr_{35}\bfr'_{13} ]^{-1}\right].
\nonumber
\end{eqnarray}
For the out-of-plane component, we notice that the second term vanishes because ${\bfr'}_{13}^{\dagger}\bfsig_y{\bfr'}_{13} = \bfs^{-1} {\bfr'}_{13}^{\dagger}(0)\,\bfs\, \bfsig_y\, \bfs^{-1}{\bfr'}_{13}(0)\,\bfs \propto \bfs^{-1} \bfp^{\da}\, \bfsig_y\, \bfp^{\da}\,\bfs = 0$.
Hence
\begin{equation}
\tr [\orbA_3^{\dagger} \bfsig_y \orbA_3 ] = \tr \left[[ \mathbf{1}- \bfr_{35}^{\dagger} {\bfr'}_{13}^{\dagger}]^{-1} \bfsig_y  [ \mathbf{1}- \bfr'_{13}\bfr_{35} ]^{-1}\right].\label{eq:A3l}
\end{equation}
Likewise, from Eq.(\ref{eq:ABr}) 
\begin{eqnarray}
\tr [\orbB_3^{\dagger} \bfsig_y \orbB_3 ] = \tr [\orbA_3^{\dagger} {\bfr}_{35}^{\dagger} \bfsig_y {\bfr}_{35} \orbA_3 ] = 0 .\label{eq:B3l}
\end{eqnarray}
A similar argument gives the results
\begin{equation}
\tr [\olbA_3^{\dagger} \bfsig_y \olbA_3 ] =0 \label{eq:A3r}
\end{equation}
\begin{equation}
\tr [\olbB_3^{\dagger} \bfsig_y \olbB_3] = \tr \left[[ \mathbf{1}- {\bfr'}_{13}^{\dagger} \bfr_{35}^{\dagger} ]^{-1} \bfsig_y  [ \mathbf{1}- \bfr_{35}\bfr'_{13} ]^{-1}\right].\label{eq:B3r}
\end{equation}
But $\bfs^T=\bfs^{-1}$, so that $\bfr_{35}(\theta)$ and $\bfr'_{13}(\theta)$ are symmetric, whereas $\bfsig_y$ is antisymmetric.
Hence
\begin{eqnarray}
& \left[[\mathbf{1} - {\bfr'}_{13}^{\dagger} \bfr_{35}^{\dagger} ]^{-1} \bfsig_y  [ \mathbf{1}- \bfr_{35}\bfr'_{13} ]^{-1}\right]^T = \cr
& \qquad \qquad -\left[[ \mathbf{1}- \bfr'_{13}\bfr_{35} ]^{-1} \bfsig_y [ \mathbf{1}- \bfr_{35}^{\dagger}{\bfr'}_{13}^{\dagger}  ]^{-1} \right],
 \nonumber
\end{eqnarray}
so that taking the trace of both sides and comparing with Eq.(\ref{eq:A3l}) we deduce that
\[\tr [\olbB_3^{\dagger} \bfsig_y \olbB_3] =-\tr [\orbA_3^{\dagger} \bfsig_y \orbA_3 ]. \]
Hence the total out-of-plane transport spin current in the spacer vanishes
\[j_y^{\text{tr}}=\orj_y - \olj_y = 0.\]

The physical reason that the out-of plane spin current vanishes in the case of exact matching can be understood as follows.
First of all, because $\bfr_{35}(0) \sim \bfr'_{35}(0) \propto \bfp^{\da}$ and $\bfr_{13}(\theta) \sim \bfr'_{13}(\theta) \propto \bfs^{-1}(\theta)\bfp^{\da}\bfs^{-1}(\theta)$, then any state reflected off the SM or PM is projected onto the down state along the same quantization axis as the magnetization. 
So it is not surprising that all states in the spacer, reflected off the PM or SM, have zero out-of -plane spin component: 
$\left\langle\bfr_{35} \bfalp'_3\right| \bfsig_y \left|\bfr_{35}(0)\bfalp_3\right\rangle =
\left\langle\bfr_{13}(\theta) \bfbet'_3\right| \bfsig_y \left|\bfr_{13}(\theta)\bfbet_3\right\rangle 
 = 0$.
This explains Eq's.~(\ref{eq:B3l}) and (\ref{eq:A3r}), and in the spacer we only have to consider right moving states originating from the left lead ($\orbA_3$) and left moving states originating from the right lead ($\olbB_3$).

Comparing Eq.'s~(\ref{eq:A3l}) and (\ref{eq:tmpa1}) we note that, as far as the out-of-plane spin current is concerned, electrons emitted from the left lead pass through the polarising magnet as if it had unit transmission matrix. Likewise for electrons from the right lead passing through the switching magnet (Eq. (\ref{eq:B3r})).  Since the contribution to out-of-plane spin current from the left and the right only involves reflections off the polarising and switching magnet interfaces, then we might reasonably expect $\orj_y = \pm \olj_y$.  Since exact matching should not lead to a vanishing of the exchange coupling, then we must have $\orj_y = \olj_y$.
\subsection{Symmetry in $\theta$}
We now examine the symmetry of the spin current with respect to the magnetization angle $\theta$ in the PM. 

We note that
\[
\bfs^{-1}(\theta)=\bfsig_z\bfs(\theta)\,\bfsig_z
\]
so that $\bfr_{13}(-\theta)=\bfs(\theta)\bfr_{13}(0)\,\bfs(-\theta)=\bfsig_z\,\bfr_{13}(\theta)\bfsig_z$, and
similarly for the other $(1,3)$ reflection and transmission matrices.
Hence from Eq.(\ref{eq:ABr}), for a system with PM magnetization $\theta_2=-\theta$
\begin{eqnarray}
\orbA_3(-\theta)&=&\left[ \mathbf{1}-\bfr'_{13}(-\theta) \bfr_{35}(0) \right]^{-1} \bft'_{13}(-\theta)\nonumber\\
&=&\left[ \mathbf{1}  -\bfsig_z\bfr'_{13}(\theta)\bfsig_z \bfr_{35}(0) \right]^{-1} \bfsig_z\bft'_{13}(\theta)\bfsig_z\nonumber\\
&=&\bfsig_z \orbA_3(\theta) \bfsig_z.\label{eq:A3sym}
\end{eqnarray}
and similarly for $\olbA_3$, $\orbB_3$ and $\olbB_3$.
Hence
\begin{eqnarray}
\tr [\olrbA_3(-\theta)^{\dagger} \bfsig_i \olrbA_3(-\theta)] = \tr [\olrbA_3(\theta)^{\dagger} \bfsig_z \bfsig_i \bfsig_z \olrbA_3(\theta)],\nonumber\\
\tr [\olrbB_3(-\theta)^{\dagger} \bfsig_i \olrbB_3(-\theta)] = \tr [\olrbB_3(\theta)^{\dagger} \bfsig_z \bfsig_i \bfsig_z \olrbB_3(\theta)].\nonumber
\end{eqnarray}
But, $\bfsig_z \bfsig_x \bfsig_z = - \bfsig_x$,  $\bfsig_z \bfsig_y \bfsig_z = - \bfsig_y$ and $\bfsig_z^3 = \bfsig_z$. 
Hence we deduce that
\begin{eqnarray}
j_x^{\text{tr}}(-\theta)=-j_x^{\text{tr}}(\theta) \  , \  j_y^{\text{tr}}(-\theta)=-j_y^{\text{tr}}(\theta) \  , \  j_z^{\text{tr}}(-\theta)=j_z^{\text{tr}}(\theta). \nonumber
\end{eqnarray}
in the spacer.
In fact it is easy to show that these symmetries also hold for spin current in the two leads.
\subsection{Reflection Symmetry}
Now let us consider a system with reflection symmetry: in which the potentials in the leads are the same $k_1=k_5$; and those in the PM and SM are the same $k_2^{\uda}=k_4^{\uda}$; and also the thickness of the PM and SM are the same.
For $\theta=0$ the system will have complete reflection symmetry, and so we expect that $t'_{13} \sim t_{35}$ etc. 
In fact this is true up to a phase:
\begin{eqnarray}
\bft'_{13}&=&e^{i(k_1-k_3)\phi}\bft_{35},\label{eq:rtid9}\\
\bft_{13}&=&e^{i(k_1-k_3)\phi}\bft'_{35},\label{eq:rtid10}\\
\bfr'_{13}&=&e^{-2i k_3 \phi}\bfr_{35},\label{eq:rtid11}\\
\bfr_{13}&=&e^{2i k_1 \phi}\bfr'_{35},\label{eq:rtid12}
\end{eqnarray}
where $\phi=y_{23}+y_{34}$.
These equations are easily derived from Eq.'s (\ref{eq:tpii1})--(\ref{eq:rii1}) at $\theta_2=\theta_4=0$.
However, as $\bfr(\theta)=\bfs^{-1}(\theta)\bfr(0)\,\bfs(\theta)$ etc., then they also hold for any $\theta_2=\theta_4$.

We might also expect that electrons incident from the left on the SM, might be equivalent to  electrons incident from the right on the PM  i.e. $\orbalp_3 \sim \olbbet_3$.
This is proved as follows.
From Eq.'s (\ref{eq:ABr}), (\ref{eq:ABl}) and (\ref{eq:rtid9})--(\ref{eq:rtid12})
\begin{eqnarray}
\olbB_3(\theta)&=&\left[ \mathbf{1} - \bfr_{35}(0)\bfr'_{13}(\theta) \right]^{-1} \bft_{35}(0)\nonumber\\
&=&\left[ \mathbf{1} - \bfr'_{13}(0)\bfs^{-1}(\theta)\bfr_{35}(0)\bfs(\theta) \right]^{-1} \bft_{35}(0)\nonumber\\
&=&e^{-i(k_1-k_3)\phi}\bfs^{-1}(\theta)\times\cr & &\hspace{0.4in} 
\left[ \mathbf{1} - \bfs(\theta)\bfr'_{13}(0)\bfs^{-1}(\theta)\bfr_{35}(0) \right]^{-1}\bfs(\theta)\bft'_{13}(0)\nonumber\\
&=&e^{-i(k_1-k_3)\phi}\bfs^{-1}(\theta)\times\cr & &\hspace{0.4in}
\left[ \mathbf{1} - \bfr'_{13}(-\theta)\bfr_{35}(0) \right]^{-1}\bft'_{13}(-\theta)\bfs(\theta)\nonumber\\
&=&e^{-i(k_1-k_3)\phi}\bfs^{-1}(\theta) \orbA_3(-\theta) \bfs(\theta), \label{eq:ref1}
\end{eqnarray}
where $\orbA_3(-\theta)$ corresponds to a system with $\theta_2=-\theta$.
Further from Eq.(\ref{eq:A3sym})
\begin{eqnarray}
\olbB_3(\theta)&=&e^{-i(k_1-k_3)\phi}\bfs^{-1}(\theta)\bfsig_3 \orbA_3(\theta) \bfsig_3  \bfs(\theta).\label{eq:B3sym5}
\end{eqnarray}
Likewise from Eq's (\ref{eq:ABr}) and (\ref{eq:ABl})
\begin{eqnarray}
\olbA_3(\theta)&=&\bfr'_{13}(\theta)\olbB_3(\theta)\nonumber\\
&=&e^{-i(k_1-k_3)\phi}\bfs^{-1}(\theta)\bfr'_{13}(0)\bfsig_3 \orbA_3(\theta) \bfsig_3 \bfs(\theta)\nonumber\\
&=&e^{-i(k_1+k_3)\phi}\bfs^{-1}(\theta) \bfsig_3 \bfr_{35}(0) \orbA_3(\theta) \bfsig_3 \bfs(\theta)\nonumber\\
&=&e^{-i(k_1+k_3)\phi}\bfs^{-1}(\theta) \bfsig_3 \orbB_3(\theta) \bfsig_3 \bfs(\theta).\label{eq:A3sym5}
\end{eqnarray}
From Eq.'s (\ref{eq:B3sym5}) and (\ref{eq:A3sym5}) we deduce
\begin{eqnarray}
\tr[\olbB_3^{\dagger}(\theta) \bfsig_i \olbB_3(\theta)] &=& \tr[\orbA_3^{\dagger}(\theta) \bfsig_3\,  \bfs(\theta)\, \bfsig_i\, \bfs^{-1}(\theta)\bfsig_3\orbA_3(\theta) ]\nonumber\\
\tr[\olbA_3^{\dagger}(\theta) \bfsig_i \olbA_3(\theta)]&=&\tr[ \orbB_3^{\dagger} (\theta) \bfsig_3\, \bfs(\theta)\, \bfsig_i\, \bfs^{-1}(\theta) \bfsig_3 \orbB_3(\theta) ],\nonumber
\end{eqnarray}
which for $\bfsig_i=\bfsig_y$ reduces to
\begin{eqnarray}
\tr[\olbB_3^{\dagger}(\theta) \bfsig_y \olbB_3(\theta)] &=& -\tr[\orbA_3^{\dagger}(\theta) \bfsig_y \orbA_3(\theta) ]\nonumber\\
\tr[\olbA_3^{\dagger}(\theta) \bfsig_y \olbA_3(\theta)] &=& -\tr[ \orbB_3^{\dagger} (\theta)  \bfsig_y \orbB_3(\theta) ].\nonumber
\end{eqnarray}
Hence the total out-of-plane transport spin current in the spacer vanishes for a symmetric system
\[j_y^{\text{tr}}=\orj_y - \olj_y = 0.\]

\begin{figure}[tp]
\begin{center}
\includegraphics[width=0.6\linewidth]{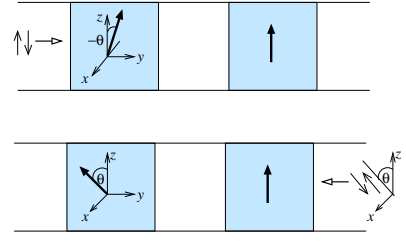}
\end{center}
\caption{\footnotesize Two systems equivalent upto a global in-plane rotation by angle $\theta$.}
\label{fig5}
\end{figure}

The root cause of why the out-of-plane spin current vanishes for a symmetric system can be understood from the preceding equations.
Firstly Eq.(\ref{eq:ref1}) (and its equivalent for $\olbA_3$) informs us that sending electrons from the left, polarized in the $z$-direction through the PM with magnetization in the $-\theta$ direction,  is equivalent to sending electrons from the right, polarized in the $\theta$-direction  through the SM.
In particular $\orbalp_3(-\theta) \sim \olbbet_3(\theta)$ and $\orbbet_3(-\theta) \sim \olbalp_3(\theta)$ upto a phase and an in-plane rotation of the entire system by $\theta$ 
 (c.f. Fig.~\ref{fig5}).
However, because the out-of-plane component of spin current $\olrj_y(\theta)$ is invariant under in-plane rotations of the entire system, and is an odd function of $\theta$, then $\olj_y(\theta)$ and $\orj_y(\theta)$ cancel in the transport spin current.

In fact we recall that we are free to choose the quantization axis of electrons emitted from the left and the right lead independently.
Hence choosing the left lead to be aligned with the PM and the right to be aligned with the SM, it is clear that in the spacer $\orj_i=\pm \olj_i$.  On physical grounds, since we do not expect the exchange coupling to vanish for a symmetric system, then we must have $\orj_y= \olj_y$.  
Furthermore, in the next section we show that if $i=0,x,z$ then $\orj_i=- \olj_i$, in any non-magnetic layer of any multilayer system.

Interestingly, the vanishing of the out-of-plane transport spin current  $J_y^{\text{tr}}$ for symmetric systems has been verified previously by numerical calculations on several realistic systems \cite{Edwards2005}, \cite{Butler2010}. 
Interestingly, this phenomenon is the reason that the authors of Ref. \cite{Butler97} originally mistakenly concluded that $J_y^{\text{tr}}$ was a quadratic function of the applied bias:  they inadvertently considered a symmetric system.

\subsection{Relation between  $\protect\orj$ and $\protect\olj$}
We now, consider a general multilayer, composed of $N$ non-magnetic (NM) and ferromagnetic (FM) layers, sandwiched consecutively so that each FM layer has NM layers either side of it. The exchange field in each FM layer is at an arbitrary angle $\theta$ to the $z$-axis in the $xz$-plane, and the potentials in each layer are arbitrary except that the potentials in the two NM leads (NM$_1$ and NM$_N$) are assumed to be equal. Schematically, we write:  $\text{NM}_1 \vert \text{FM}_2(\theta_2) \vert \text{NM}_3 \vert \text{FM}_4(\theta_4) \vert \ldots \vert \text{FM}_{N-1}(\theta_{N-1}) \vert \text{NM}_N $.

In the left lead (layer $1$), we have 
\[\orj_i^{(1)}=\tr(\bfsig_i - \bfr_{1,N}^{\dag}\bfsig_i \bfr_{1,N})=\tr(\bfsig_i \bft_{1,N}\bft_{1,N}^{\dag}),\]
where we have used Eq.~(\ref{eq:rtid2a}),  ${\bft}_{1,N} \bft^{\dag}_{1,N} + {\bfr}_{1,N} {\bfr}_{1,N}^{\dagger}=\mathbf{1}$. Clearly then $\orj_i^{(1)}=-\olj_i^{(1)}=-\tr(\bft_{1,N}^{\dag}\bfsig_i \bft_{1,N})$.  In exactly the same way, see that 
\[\olj_i^{(N)}=\tr(\bfr_{1,N}^{' \dag}\bfsig_i \bfr'_{1,N} - \bfsig_i)=-\tr(\bfsig_i \bft'_{1,N}\bft_{1,N}^{' \dag})=-\olj_i^{(N)}.\]
Hence we deduce that there are no components of the exchange coupling in the leads.

To deduce the relationship between $\orj$ and $\olj$ in a general non-magnetic layer we need to proceed more carefully.
In any conducting non-magnetic layer $n$, the spin current from electrons of spin $\nu=\ua,\da$  incident on the left lead is given by (Equations (\ref{eq:T12}) and (\ref{eq:sc1}))
\begin{eqnarray}
\orj_i^{(n)\nu} =  k_n \orbPsi_n^{\nu \dag} \bfSig_i \orbPsi_n^{\nu} =  k_n \orbPsi_N^{\nu \dag}\bfT_{nN}^{\dag} \bfSig_i \bfT_{nN} \orbPsi_N^{\nu} \nonumber
\end{eqnarray}
where
\begin{eqnarray}
\bfSig_i = \begin{pmatrix}
\bfsig_i & \mathbf{0} \\\mathbf{0} &-\bfsig_i
\end{pmatrix} .\nonumber
\end{eqnarray}
However since $\orbPsi_N^{\nu}=\begin{pmatrix}
\bft'_{1N}\bfalp_1^{\nu} \\ \mathbf{0}
\end{pmatrix}$
then the total spin current $j=j^{\ua}+j^{\da}$ incident from the left is given by
\begin{eqnarray}
\orj_i^{(n)} = k_n \tr \left[\begin{pmatrix}
\bft^{'\dag}_{1N}& \mathbf{0}
\end{pmatrix}    (\bfT_{nN}^{\dag} \bfSig_i \bfT_{nN}) \begin{pmatrix}
\bft'_{1N}\\ \mathbf{0}
\end{pmatrix} \right].\nonumber
\end{eqnarray}
Likewise, for electrons of spin $\nu=\ua,\da$ incident from the right, the spin current in layer $n$ is
\begin{eqnarray}
\olj_i^{(n)\nu} =  k_n \olbPsi_n^{\nu \dag} \bfSig_i \olbPsi_n^{\nu} =  k_n \olbPsi_N^{\nu \dag}\bfT_{nN}^{\dag} \bfSig_i \bfT_{nN} \olbPsi_N^{\nu}. \nonumber
\end{eqnarray}
Now, $\olbPsi_N^{\nu}=\begin{pmatrix}
\bfr'_{1N}\bfalp^{\nu}_N \\ \bfalp^{\nu}_N
\end{pmatrix}$, so that the total spin current $j=j^{\ua}+j^{\da}$ incident from the right is 
\begin{eqnarray}
\olj_i^{(n)} = k_n \tr \left[\begin{pmatrix}
\bfr'^{\dag}_{1N}& \mathbf{1}
\end{pmatrix}    (\bfT_{nN}^{\dag} \bfSig_i \bfT_{nN}) \begin{pmatrix}
\bfr'_{1N}\\ \mathbf{1}
\end{pmatrix} \right]. \nonumber
\end{eqnarray}
Using Equation~(\ref{eq:rtid2b})  ${\bft'}_{1,N} \bft^{' \dag}_{1,N} + {\bfr'}_{1,N} {\bfr'}_{1,N}^{\dagger}=\mathbf{1}$, this becomes
\begin{eqnarray}
\olj_i^{(n)} &=& -\orj_i^{(n)}  + \nonumber\\& & \hspace{0.2in} k_n \tr \left[[\bfL_i]_{11} + [\bfL_i]_{22} + \bfr{' \dag}_{1N}[\bfL_i]_{12} + [\bfL_i]_{21}\bfr'_{1N} \right]\nonumber\\ 
&=& -\orj_i^{(n)}  +  k_n \tr[ \bfW \bfL_i]  \nonumber
\end{eqnarray}
where $\bfL_i = \bfT_{nN}^{\dag} \bfSig_i \bfT_{nN}$ and  $\bfW = \begin{pmatrix}
\mathbf{1} & \bfr'_{1N}\\ {\bfr'}^{\dag}_{1N} & \mathbf{1}
\end{pmatrix}$.

In the case of charge current, $i=0$ and so by Eq.~(\ref{eq:Tinv}), $\bfL_0=k_N k_n^{-1} \bfSig_0$. It follows that $\tr[ \bfW \bfL_0]=0$ and hence for charge current $\olj_0^{(n)} = -\orj_0^{(n)}$.

For spin current, we proceed as follows.
Clearly both $\bfL_i$ and $\bf W$ are hermitian.
Further, from Equation~(\ref{eq:Tconj2}), if $i=x$ or $z$ then ${\bfL}^*_i=-\bfI \bfL_i \bfI$, where $\bfI=\begin{pmatrix}\mathbf{0} & \mathbf{1} \\ \mathbf{1} & \mathbf{0} \end{pmatrix}$.  
Further, because $\bfr'$ is symmetric (Eq.~\ref{eq:rtrans}), then $\bfW^*=\bfI \bfW \bfI$.  
So on the one hand we get
\[
(\tr [\bfW \bfL_i])^* = \tr [\bfW \bfL_i]^{\dag} = \tr [\bfW \bfL_i],
\]
whilst on the other hand we get
\[
(\tr [\bfW \bfL_i])^* = \tr [\bfW \bfL_i]^{*} = -\tr [\bfW \bfL_i].
\]
So for $i=x$ or $z$ we conclude that $\tr [\bfW \bfL_i] = 0$, and hence
\begin{eqnarray}
\olj_i^{(n)} &=& -\orj_i^{(n)} \quad \text{for}\quad i=0,x,z.\label{eq:lrscequiv}
\end{eqnarray}

Hence the only component of the exchange coupling is the out-of-plane component.
\section{Asymptotic Properties of Spin Current in the Spacer}
\label{sec:asymptotic}
In this section we apply our formalism to examine the oscillatory behaviour of the integrated spin current, as the thickness of the spacer is increased. A similar calculation has been previously performed for the electrical conductance in a parabolic band \cite{PhysRevB.52.R6983}. There it was shown that when all contributions to transmission are included, by integrating over the in-plane momenta $\bkp$, the resulting expression oscillates as a function of the spacer thickness $L$, with the amplitude decreasing as $1/L$.
This oscillation is RKKY-like and arises from near those $\bkp$-points where the spacer Fermi surface has extrema in the growth direction ($\bkp=\mathbf{0}$).
 Furthermore, when the potential profile is a rectangular well, with the Fermi energy level near the top, another oscillation is observed. This second period decays as $L^{-3/2}$ and arises from the boundary effects, near the top of the well, where transmission vanishes non-analytically. \par Here we perform a similar analysis for all components of the spin current, considering the cases where the potential profile of the multilayer is a double barrier or a double well, that is, where the potentials in the magnets are greater or less than those in the leads and the spacer, respectively. In each case we find only RKKY type periods. This is because (after switching to polar coordinates) the integrand of the conductance has leading order $\Gamma \sim O(k^{-1/2})$ while the integrand of spin current components is $j_i \sim O(k^{1/2})$, where $k$ is the the out-of-plane wave vector. So the spin current density tends to zero in a smooth way near the zone boundary and the total spin current does not exhibit the non-RKKY-like period.
Following the discussion in Section \ref{sec:origin} spin current components in the spacer are obtained by integrating spin current density over the permissible values of $\bkp$, at the Fermi energy $E=E_{\text F}$. Since we assume the system having rotational symmetry in momentum space, we can switch to polar coordinates in $xz$-plane whereby the total current is given by the following formula
\begin{equation}
J_{i}(L)=2\pi\int_{0}^{k_{\text{F}}}
j_{i}(\bs{k}_{\parallel},L)k_{\parallel}dk_{\parallel},\label{eq:total_current_int}
\end{equation}
where $k_{\text{F}} = \sqrt{2m(E_{F}-V)}/\hbar$, $V=V_1=V_3=V_5$ is the potential in the spacer and the leads, and $k_{\parallel}=|\bs{k}_{\parallel}|$. The double barrier (well) profile is therefore characterised by the condition $V_i-\Delta / 2 > V$ ($V_i+\Delta / 2 < V$) where $i=2,4$. Assuming that reflections off the magnets are not too strong ($||\bfr'_{13}\bfr_{35}||\ll1$) we retain only the first-order reflections in the series expansion of the amplitude. Here we denote $k\equiv k_{3}$ for the out-of-plane wave-vector in the spacer and also suppress the layer index at the current density $j_{i}$, because we are only interested in the spacer current in this section. For the current generated by electrons incident from the left we obtain
\begin{equation*}
\begin{aligned}\overrightarrow{j}_{i} &=k\,\tr{\left\{\left(\overrightarrow{\mathbf{a}}\overrightarrow{\mathbf{a}}^{\dagger}-\overrightarrow{\mathbf{b}}\overrightarrow{\mathbf{b}}^{\dagger}\right)\bfsig_{i} \right\}}\\
&=k\,\tr{\left\{\left(\overrightarrow{\mathbf{a}}\overrightarrow{\mathbf{a}}^{\dagger}-\bfr_{35}\overrightarrow{\mathbf{a}}\overrightarrow{\mathbf{a}}^{\dagger}\bfr_{35}^{\dagger}\right)\bfsig_{i} \right\}}.
\end{aligned}
\end{equation*}
Expanding $\overrightarrow{\mathbf{a}}$ given by Eq.\eqref{eq:ABr}, and retaining one reflection term we obtain
\begin{equation*}
\overrightarrow{\mathbf{a}}\approx\bft'_{13}+\bfr'_{13}\bfr_{35}\bft'_{13}.
\end{equation*}
We are interested in the terms that are periodic in $L$.
This periodicity is contained only in $\bfr_{35}$ via the factor of $e^{2ik\left(L_{\text{PM}}+L\right)}$. Hence, after collecting only the terms containing $\bfr_{35}$, the periodic part of the right moving current, $\oscr{j}_{i}$, becomes
\begin{equation*}
\oscr{j}_{i} =k\,\tr{\left\{ \left(\oscr{\bfp}+\overrightarrow{\bfp}^{\sim\dagger}\right)\bfsig_{i}\right\}}=2k\Re{\tr{\left\{\oscr{\bfp}\bfsig_{i}\right\}}},
\end{equation*}
where
\begin{equation*}
\oscr{\bfp}=\bfr_{13}^{'}\bfr_{35}\bft_{13}^{'}\bft_{13}^{'\dagger} - \bfr_{35}\bfr_{13}^{'}\bfr_{35}\bft_{13}^{'}\bft_{13}^{'\dagger}\bfr_{35}^{\dagger}, 
\end{equation*}%
The remaining terms all contain equal number of occurencies of $\bfr_{35}$ and $\bfr^{\dagger}_{35}$, so do not depend on $L$. Therefore, they only contribute to the non-oscillatory constant background part of the current. Now, $\bfrprime_{13}$ contributes a phase shift of $e^{-2ikL_{\text{PM}}}$, so the resulting periodic factor of $\oscr{\bfp}$ is $e^{2ikL}$. We can therefore write
\begin{equation*}
\oscr{\bfp} = \overrightarrow{{\boldsymbol{\rho}}}e^{2ik L},
\end{equation*}
and define amplitude $\overrightarrow{A}_{i}=2k\,\tr{\left\{\overrightarrow{{\boldsymbol{\rho}}}\bfsig_{i}\right\} }$.
Repeating the calculation for the left-moving current where
\begin{equation*}
\begin{aligned}\overleftarrow{j}_{i} &=k\,\tr{\left\{\left(\bfr'_{13}\olbB\ \olbB^{\dagger}\bfr^{'\dagger}_{13}-\olbB\olbB^{\dagger}\right)\bfsig_{i} \right\}},
\end{aligned}
\end{equation*}
and
\begin{equation*}
\olbB \approx \bft_{35} + \bfr_{35}\bfr'_{13}\bft_{35},
\end{equation*}
we eventually obtain the total amplitude $A_i(k_{\parallel}) = \overrightarrow{A}_i(k_{\parallel}) + \overleftarrow{A}_i(k_{\parallel})$ (up to one reflection) and the following expresson for the oscillatory part of the current density 
\begin{equation}
j^{\sim}_{i} = 
\Re\left[A_{i}(k_{\parallel})e^{2ik(k_{\parallel})L}\right].
\label{eq:current_trace_amplitude}
\end{equation}
Eq.\eqref{eq:total_current_int} is now evaluated using the stationary phase approximation to derive an asymptotic formula for the oscillatory part of the spin current, valid for large values of $L$. Since we are considering the parabolic band model where $k(\kpar)$ is only stationary at $\kpar=0$, we obtain :
\begin{equation}
J^{\sim}_{i}(L) =\Re\frac{\pi A_{i}(0)k(0)}{iL}e^{2ik(0)L}+O\left(\frac{1}{L^{2}}\right).
\label{eq:stat_phase}
\end{equation}
In doing so, we have replaced the upper limit of integration with $\infty$, which is justified by the fact that the integrand vanishes identically beyond the top of the barrier (bottom of the well).
\begin{figure}[tp]
	\begin{center}
		\includegraphics[width=1.0\linewidth]{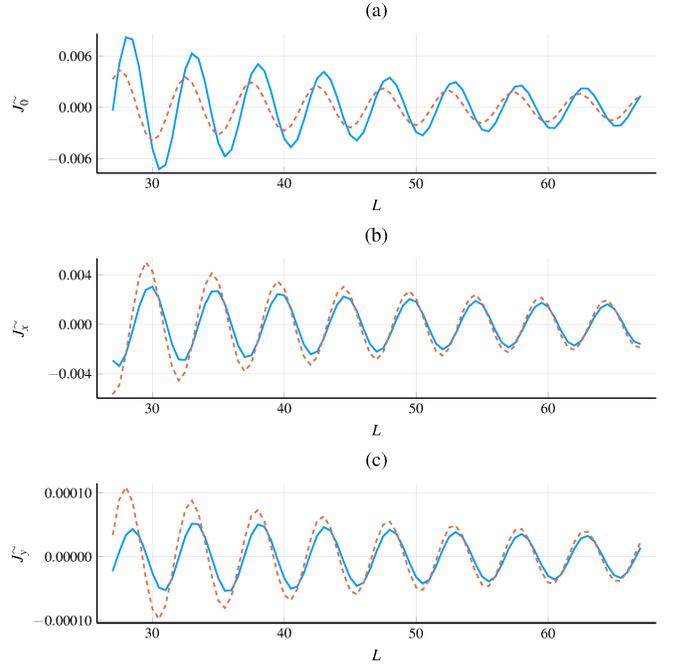}
	\end{center}
	\caption{\footnotesize Double barrier. Plots of the full numeric calculation (solid line) and the asymptotic approximation (dashed line) of (a) $J^{\sim}_0$ (here we set $2e/\hbar=1$), (b), $J^{\sim}_x$ and (c) $J^{\sim}_y$ as functions of spacer thickness.}
	\label{fig7}
\end{figure}
\begin{figure}[tp]
	\begin{center}
		\includegraphics[width=1.0\linewidth]{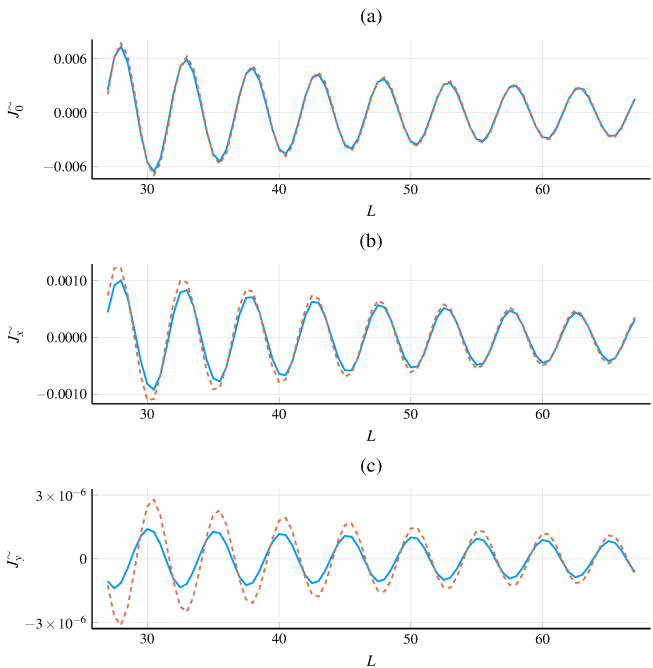}
	\end{center}
	\caption{\footnotesize  Double well. Plots of the full numeric calculation (solid line) and the asymptotic approximation (dashed line) of (a) $J^{\sim}_0$ (here we set $2e/\hbar=1$), (b), $J^{\sim}_x$ and (c) $J^{\sim}_y$ as functions of spacer thickness.}
	\label{fig8}
\end{figure}%

To check the accuracy of this approximation, we compared it with a full numerical calculation for both the double-well and barrier cases.
We considered a model with relatively shallow potentials, so that confinement is not too strong, and the single-reflection approximation holds well. Potentials in the spacer, leads and magnets were chosen as in Table \ref{tab:device}, where positive and negative potentials in the magnets correspond to the double-barrier and double-well profiles, respectively, together with a Fermi energy level set to $E_{\text{F}}=0.4$.
\begin{table}[ht]
	\begin{tabular}{|c|c|c|c|c|}
		\hline 
		& $V$ & $\Delta$ & $\theta$ & $y_{n+1}-y_{n}$\tabularnewline
		\hline 
		Lead 1 & 0.0 & 0.0 & 0.0 & -\tabularnewline
		\hline 
		PM & $\pm0.5$ & 0.05 & 0.6 & 7.0\tabularnewline
		\hline 
		Spacer & 0.0 & 0.0 & 0.0 & 20.0\tabularnewline
		\hline 
		SM & $\pm0.5$ & 0.05 & 0.0 & 3.0\tabularnewline
		\hline 
		Lead 2 & 0.0 & 0.0 & 0.0 & -\tabularnewline
		\hline 
	\end{tabular}
	\caption{Device parameters used to obtain figures \ref{fig7} and \ref{fig8}}.
	\label{tab:device}
\end{table}
Our results are shown in figures \ref{fig7} and \ref{fig8}, where we plot the numerically integrated current Eq.\eqref{eq:total_current_int} as a function of the spacer thickness and compare it with the asymptotic approximation derived in Eq.\eqref{eq:stat_phase}, for the cases of the barrier and well, respectively. 
We observe excellent agreement between the asymptotic approximation and the full numerical results in all cases.
Hence we conclude that all components of the spin current oscillate with spacer thickness, with a period determined by the Fermi surface extrema in the growth direction and with an asymptotic decay given by the inverse of the spacer thickness.

\section{Conclusion}
In this communication we have developed the Landauer method and transfer matrix formalism to deal with spin currents in the ballistic regime for a simple one electron parabolic band model.
The formalism provides an elegant and physically transparent description of spin currents which allows closed form expressions to be obtained in terms of the reflection and transmission matrices of the system.
In turn, the algebraic properties of the reflection and transmission matrices can be used to derive analytic results concerning spin currents.
We have used the formalism to understand the origin and symmetries of the components of the spin current for the standard polarizing-magnet/non-magnetic spacer/switching-magnet geometry.
In  particular, within this formalism we have: 
shown that the out-of-plane spin current in the spacer arises as a consequence of reflections of carriers between the polarizing and switching magnets;
explained why the bias dependent part of the out-of-plane spin current vanishes in the spacer if the multilayer is symmetric or there is exact matching of potentials of one spin band across the multilayer; 
explained why only the out-of-plane component survives in the absence of bias.
In conjunction with the Landauer formalism, we have also applied the stationary phase approximation to the calculate the components of the total spin current, accurate in the limit of relatively weak reflection and large spacer thickness. 
We have shown that in this limit, the oscillatory parts of all components of the spin current (including the charge current), can be accurately described by the stationary phase approximation, in the case of both a potential well and potential barrier i.e. where the potentials of the magnets lie below or above the lead and spacer respectively. 
The oscillation period is given by those $\bkp$-points where the spacer Fermi surface has extrema in the growth direction, and the oscillation decay is the inverse of the spacer thickness.  
Such approximations allow us to understand the behaviour of the spin current and could be of considerable use if extended to multi-orbital models where accurate numerical calculation of spin current is very challenging.
\section{Appendix: Symmetry Properties of the transfer matrix}
In this appendix we derive some symmetry properties of the transfer matrix and of the reflection and transmission matrices.
Throughout this section, we consider a general multilayer, composed of $N$ non-magnetic (NM) and ferromagnetic (FM) layers, sandwiched consecutively so that each FM layer has NM layers either side of it. The exchange field in each FM layer is at an arbitrary angle $\theta$ to the $z$-axis in the $xz$-plane, and the potentials in each layer are arbitrary except that the potentials in the two NM leads (NM$_1$ and NM$_N$) are assumed to be equal. Schematically, we write:  $\text{NM}_1 \vert \text{FM}_2(\theta_2) \vert \text{NM}_3 \vert \text{FM}_4(\theta_4) \vert \ldots \vert \text{FM}_{N-1}(\theta_{N-1}) \vert \text{NM}_N $.
\subsection{Symmetry of reflection and transmission matrices under transposition}
First consider the transmission matrix $t_{n,n+1}$ at $\theta_n=0$ or $\theta_{n+1}=0$ . From Equations~(\ref{eq:tpii1}) and (\ref{eq:tii1}), we get
\[
\left. \bft_{n,n+1}\right|_{\substack{
\theta_n=0\\
\theta_{n+1}=0}}=\left. \bfk_{n+1}\bfk_n^{-1}\,\bft'_{n,n+1}\right|_{\substack{
\theta_n=0\\
\theta_{n+1}=0}},
\]
where $\bfk_n=\diag[k_n^{\ua},k_n^{\da}]$.

Now let us temporarily set $\theta_2=0$ and consider $\bft_{13}$ and  $\bft'_{13}$. We have
\begin{eqnarray}
\bft_{13} = \bft_{12} (\mathbf{1} - \bfr_{23}\bfr'_{12})^{-1} \bft_{23} \nonumber\\
\bft'_{13} = \bft'_{23} (\mathbf{1} - \bfr'_{12}\bfr_{23})^{-1} \bft'_{12}  , \nonumber
\end{eqnarray}
hence taking the transpose we get $\bft'_{13}(0)^T= k_1 k_3^{-1} \bft_{13}(0) $ because everything is diagonal.
However, by Equation~(\ref{eq:thdep}), $\bft_{13}'(\theta_2)={\bf s}^{-1}(\theta_2).\bft'_{13}(0).{\bf s}(\theta_2)$ and $\bfs^T=\bfs^{-1}$, so we deduce that $\bft'_{13}(\theta_2)^T =  k_1 k_3^{-1} \bft_{13}(\theta_2) $.  Similarly, we get
\begin{equation}
\bft'_{n,n+2}(\theta_{n+1})^T =  k_n k_{n+2}^{-1} \bft_{n,n+2}(\theta_{n+1}) ,\label{eq:apxttp2}
\end{equation}
where layers $n$ and $n+2$ are non-magnetic.
By Equation~(\ref{eq:thdep}), we also note that $\bft'_{n,n+2}$, $\bft_{n,n+2}$, $\bfr'_{n,n+2}$, and $\bfr_{n,n+2}$, are all symmetric.

Now consider reflection and transition matrices between layers 1 and 5, for general values of $\theta_2$ and $\theta_4$. 
We have
\[
\bfr_{15} = \bfr_{13}  +  \bft_{13} \bfr_{35} (\mathbf{1} - \bfr'_{13}\bfr_{35})^{-1} \bft'_{13}
\]
hence
\[
\bfr_{15}^T = \bfr_{13}  +   \bft'_{13} (\mathbf{1} - \bfr_{35} \bfr'_{13})^{-1} \bfr_{35}  \bft_{13} 
= \bfr_{15}
\]
where we have used Equation~(\ref{eq:apxttp2}).
Similarly we deduce that
\[
\bfr_{n,n+4}^T = \bfr_{n,n+4}  \quad, \quad {\bfr'}^T_{n,n+4} = \bfr'_{n,n+4}
\]
where layers $n$ and $n+4$ are non-magnetic.  For the transmission matrices we have
\begin{eqnarray}
\bft'_{15} = \bft'_{35} (\mathbf{1} - \bfr'_{13}\bfr_{35})^{-1} \bft'_{13} \nonumber\\
\bft_{15} = \bft_{13} (\mathbf{1} - \bfr_{35}\bfr'_{13})^{-1} \bft_{35} \nonumber
\end{eqnarray}
so that
\[
{\bft'}^T_{15} = {\bft'}^T_{13} (\mathbf{1} - \bfr_{35}\bfr'_{13})^{-1} {\bft'}^T_{35} = k_1 k_5^{-1 }\bft_{15}.
\]
More generally, we obtain ${\bft'}^T_{n,n+4} = k_n k_{n+4}^{-1 }\bft_{n,n+4}$ where layers $n$ and $n+4$ are non-magnetic.

Proceeding in this way we can deduce that
\begin{eqnarray}
{\bfr'}^T_{n,m} &=&\bfr'_{n,m}\quad ,\quad\bfr_{n,m}^T =\bfr_{n,m}  \label{eq:rtrans}\\
{\bft'}^T_{n,m} &=&  k_n k_m^{-1 }\bft_{nm}. \label{eq:ttrans}
\end{eqnarray}
when $n$ and $m$ are non-magnetic.

\subsection{Symmetry of the transfer matrix under complex conjugation}
If $k^{\ua}$ and $k^{\da}$ are both real, then from Equation~(\ref{eq:X}) we have
\[
\bfX(k^{\uda})^*=\bfX(k^{\uda}).\bfI \quad \text{where} \quad 
\bfI = \begin{pmatrix}
\mathbf{0} & \mathbf{1} \\ \mathbf{1} & \mathbf{0}
\end{pmatrix}.
\]
If $k^{\ua}$ and $k^{\da}$ are both pure imaginary, then $\bf X$ is real, and hence $\bfX(k^{\uda})^*=\bfX(k^{\uda})$.

If one of $k^{\ua}$, $k^{\da}$ is real and one is pure imaginary, then since both $\bfe$ and $\bfk$ are diagonal we get a mixture of the above two cases, giving
\[
\bfX(k^{\uda})^*=\bfX(k^{\uda}).\bfJ(k^{\uda}) \quad \text{where} \quad 
\bfJ^2 = \mathbf{1},
\]
and $\bfJ$ is explicitly given by 
$\bfJ(k^{\uda}) = \begin{pmatrix}
\mathbf{1}-\bfj & \ \bfj \\  \bfj & \mathbf{1}-\bfj 
\end{pmatrix}$,
where $\bfj = \diag[f(k^{\ua}),f(k^{\da})]$, and $f(k)=\Re (k) / k = 1$ or $0$ if $k$ is pure real or imaginary respectively.

Hence the transfer matrix between adjacent layers satisfies $\bfT_{n,n+1}^*=\bfJ(k^{\uda}_n) \bfT_{n,n+1} \bfJ(k^{\uda}_{n+1})$, from which we deduce that the transfer matrix between two general layers satisfies
\[
\bfT_{n,m}^*=\bfJ(k^{\uda}_n) \bfT_{n,m} \bfJ(k^{\uda}_{m}).
\]
In particular, if layers $n$ and $m$ are conducting, then
\[
\bfT_{n,m}^*=\bfI\, \bfT_{n,m}\, \bfI,\label{eq:Tconj2}
\]
so that $\bfT$ must have the form
\[
\bfT_{n,m}=\begin{pmatrix}
\bftau_{11} & \bftau_{12} \\ \bftau_{12}^* & \bftau_{11}^*
\end{pmatrix}.
\]
%

\begin{acknowledgments}
AU wishes to thank J. Mathon and T. Human for fruitful discussions.
\end{acknowledgments}

%

\end{document}